\documentclass[useAMS,usenatbib]{mn2e}
\usepackage{epsfig}

\title[First Stars: Growth Under Feedback]
      {The First Stars: Mass Growth Under Protostellar Feedback}

\author[A. Stacy, T. H. Greif and V. Bromm]
       {Athena Stacy$^{1,2}$\thanks{E-mail: minerva@astro.as.utexas.edu}, Thomas H. Greif $^{3}$ and Volker Bromm$^{1,2,3}$\\
 $^{1}$Department of Astronomy, University of Texas, Austin, TX 78712, USA \\
 $^{2}$Texas Cosmology Center, University of Texas, Austin, TX 78712, USA \\
 $^{3}$Max-Planck-Institut f\"{u}r Astrophysik, Karl-Schwarzschild-Str. 1, 85741 Garching, Germany }

\pagerange{\pageref{firstpage}--\pageref{lastpage}}
\pubyear{2011}

\begin{document}

\maketitle
\topmargin-1cm

\label{firstpage}

\begin{abstract}
We perform three-dimensional cosmological simulations to examine the growth of metal-free, Population III (Pop III) stars under radiative feedback.  We begin our simulation at $z=100$ and trace the evolution of gas and dark matter until the formation of the first minihalo.  We then follow the collapse of the gas within the minihalo up to densities of $n = 10^{12}$ cm$^{-3}$, at which point we replace the high-density particles with a sink particle to represent the growing protostar.  We model the effect of Lyman-Werner (LW) radiation emitted by the protostar, and employ a ray-tracing scheme to follow the growth of the surrounding H~{\sc ii} region over the next 5000 yr.  We find that a disk assembles around the first protostar, and that radiative feedback will not prevent further fragmentation of the disk to form multiple Pop III stars.  Ionization of neutral hydrogen and photodissociation of H$_2$ by LW radiation leads to heating of the dense gas to several thousand Kelvin, and this warm region expands outward at the gas sound speed.  Once the extent of this warm region becomes equivalent to the size of the disk, the disk mass declines while the accretion rate onto the protostars is reduced by an order of magnitude.  
This occurs when the largest sink has grown to $\sim 20$~M$_{\odot}$ while the second sink has grown to $\sim 7$~M$_{\odot}$,  
and we estimate the main sink will approach an asymptotic value of 30~M$_{\odot}$ by the time it reaches the main sequence. 
Our simulation thus indicates that the most likely outcome is a massive Pop III binary.  However, we simulate only one minihalo, and the statistical variation between minihaloes may be substantial.  If Pop III stars were typically unable to grow to more than a few tens of solar masses, this would have important consequences for the occurence of pair-instability supernovae in the early Universe as well as the Pop III chemical signature in the oldest stars observable today.        

\end{abstract}


\section{Introduction}

The first stars were the earliest luminous objects to form at the end of the `Dark Ages' that followed the emission of the Cosmic Microwave Background.  These stars are thought to have formed around $z\ga 20$ within minihaloes of mass $M\sim 10^6$\,M$_{\odot}$, when the dark matter (DM) potential well of the minihalo becomes large enough to gather in primordial gas
(e.g. \citealt{haimanetal1996,tegmarketal1997,yahs2003}).  
Also known as Population III (Pop III) stars, they are the early drivers of cosmic evolution (e.g. \citealt{barkana&loeb2001,bromm&larson2004,ciardi&ferrara2005,glover2005,byhm2009,loeb2010}).  Through their emission of ionizing radiation over their lifetime, Pop III stars are responsible for the beginning of cosmic reionization 
(e.g. \citealt{kitayamaetal2004,syahs2004,whalenetal2004,alvarezetal2006,johnsongreif&bromm2007}).  With the release of the first metals through their possible supernova (SN) deaths, they also provided the early metal enrichment of the intergalactic medium (IGM; \citealt{madauferrara&rees2001,moriferrara&madau2002,brommyoshida&hernquist2003,wada&venkatesan2003,normanetal2004,tfs2007,greifetal2007,greifetal2010,wise&abel2008,maioetal2011}; recently reviewed in \citealt{karlssonetal2011}).

The extent to which Pop III stars can modify their surroundings is crucially dependent upon their mass, as this is the main characteristic that determines the star's luminosity and ionizing radiation output.  Furthermore, their mass determines the type of stellar death they will undergo (\citealt{hegeretal2003}).  Only
stars in the mass range of 140~M$_{\odot}$~$<$~$M_{*}$~$<$~260~M$_{\odot}$ are predicted to explode as pair-instability supernovae (PISNe; \citealt{heger&woosley2002}), 
while stars with masses in the range 40~M$_{\odot}$~$<$~$M_{*}$~$<$~140~M$_{\odot}$ are thought to collapse directly into black holes.
Below 40 M$_{\odot}$, stars are again expected to explode as core-collapse SNe, leaving behind a neutron star or black hole. 

Previous work has found that Pop III stars begin as very small protostars of initial mass $\sim 5 \times 10^{-3}$ M$_{\odot}$ (\citealt{omukai&nishi1998,yoh2008}).  
Continued accretion onto the protostar over time ultimately leads to stars significantly larger than the initial seeds (e.g. \citealt{omukai&palla2003,bromm&loeb2004}).  
The final mass reached by Pop III stars therefore depends on the rate and duration of the accretion.  
Early numerical studies found that Pop III stars are likely to reach very high masses ($\ga 100$ M$_{\odot}$; e.g. \citealt{abeletal2002,brommetal2002}).  The lack of metal and dust cooling in primordial gas leads to higher temperatures and greater accretion rates as compared to current star-forming regions such as the giant molecular clouds within the Milky Way.  

The accretion history of Pop III stars depends critically on the radiative feedback exerted during their growth phase, an effect not included in the earliest three-dimensional simulations.  
The strength of protostellar feedback in turn hinges upon the three-dimensional structure of the surrounding gas.  For instance,  \cite{omukai&inutsuka2002} find that for spherically symmetric accretion onto an ionizing star, the formation of an H~{\sc ii} region in fact does not impose an upper mass limit to Pop III stars.
Similarly, the analytical study by \cite{mckee&tan2008} found that Pop III stars can grow to greater than 100 M$_{\odot}$ even as a protostar's own radiation ionizes its surroundings.  Although the ionization front will expand along the polar regions perpendicular to the disk, mass from the disk itself can continue to accrete onto the star until this is halted through photoevaporation.  Recent studies of present-day star formation also support the picture that stars can reach very high masses through disk accretion.  For instance, numerical studies by \cite{krumholzetal2009} and \cite{kuiperetal2011} both find that strong feedback from radiation pressure will not halt mass flow through the stellar disk before the star has reached $\ga 10$  M$_{\odot}$.  The final stellar masses are likely even greater, though the simulations were not followed for sufficiently long to determine this.  There is also growing observational evidence that massive star-forming regions exhibit disk structure and rotational motion 
(e.g. \citealt{cesaronietal2007,beutheretal2009,liuetal2011}).

Meanwhile, the picture of a single massive Pop III star forming in a minihalo has been complicated by more recent work.  Simulations by Clark et al. (2008, 2011a)\nocite{clarketal2008,clarketal2011a} employing idealized initial conditions found that primordial star-forming gas can undergo fragmentation to form Pop III multiple systems, while the simulations of \cite{turketal2009} and \cite{stacyetal2010} established such fragmentation also when initialized on cosmological scales.  Further work revealed that gas fragmentation can occur even on very small scales ($\sim$ 10 AU) and in the majority of minihaloes, if not nearly all (Clark et al. 2011b, \nocite{clarketal2011b} \citealt{greifetal2011}).  These studies tentatively imply that the typical Pop III mass may be somewhat lower than $\sim 100$ M$_{\odot}$. \cite{smithetal2011} recently found that, even under feedback from protostellar accretion luminosity, such fragmentation may be reduced but not halted.  While \cite{smithetal2011} included a heating term derived from the accretion luminosity, they did not explicitly account for molecular photodissocation by Lyman-Werner (LW) radiation from the protostar, and they did not include the effects of ionizing radiation that will become important once the stars have grown to larger masses ($\ga 10\, \rm M_{\odot}$).  
The recent two-dimensional calculation by \cite{hosokawaetal2011} found that Pop III stars will grow to 40 M$_{\odot}$,  after which accretion will be shut off by radiative feedback.  However, the effects of three-dimensional non-axisymmetry could not be addressed in this study.

Whether Pop III stars can attain very large masses under feedback and while within a multiple system is pivotal to understanding their potential for large-scale feedback effects, such as the suppresion or enhancement of the star-formation rate in neighboring minihaloes.  It also determines whether they may be observed as gamma-ray bursts (GRBs) or extremely energetic PISNe (e.g. \citealt{bromm&loeb2002,bromm&loeb2006,gouetal2004,belczynskietal2007,stacyetal2011}).   Furthermore, the fragmentation seen in recent work, along with the possible ejection of low-mass Pop III stars from their host star-forming disks (e.g. \citealt{greifetal2011, smithetal2011}), opens the possibility that small, long-lived Pop III stars may still be observed today.  This depends, however, on uncertain factors such as the final masses reached by the ejected stars and the amount of metal-enriched material accreted at later times while being incorporated into larger galaxies, which could mask the stars as Pop II (e.g. \citealt{frebeletal2009,johnson&khochfar2011}).  

To further explore the range of masses possible for Pop III stars, we perform a three-dimensional cosmological simulation to study the feedback effects of a protostar on its own accretion and on further fragmentation within its host minihalo.  We initialize the simulation with sufficient resolution to follow the evolution of the star-forming gas up to densities of 10$^{12}$ cm$^{-3}$.  At this density we employ the sink particle method, allowing us to study the subsequent disk formation and fragmentation of the gas over the following $\sim$ 5000 yr.  We include H$_2$ dissociating LW feedback from the most massive star in the simulation, and we use a ray-tracing scheme to follow the growth of the star's H~{\sc ii} region once it has become massive enough to ionize the surrounding gas.  We compare this to a simulation with the same initialization but no radiative feedback.  This allows for a direct evaluation of how radiative feedback alters the mass growth of Pop III stars, the rate of disk fragmentation, and the formation of additional stars within the disk.  We describe our numerical methodology in Section 2, while in Section 3 we present our results.  We conclude in Section 4.

\section{Numerical Methodology}

\subsection{Initial Setup}
We ran our simulation using GADGET 2, a widely-tested three-dimensional N-body and SPH code (\citealt{springel2005}). We initialized our simulation using a snapshot from the previous simulation of \cite{stacyetal2010}.  In particular, we chose the snapshot in which the dense gas in the center of the minihalo has first reached $10^8$ cm$^{-3}$.  This simulation was originally initialized at $z=100$ in a periodic box of length 140 kpc (comoving) using both SPH and DM
particles.  This was done in accordance with a $\Lambda$CDM cosmology
with $\Omega_{\Lambda}=0.7$, $\Omega_{\rm M}=0.3$, $\Omega_{\rm B}=0.04$, and $H_0=70$ km s$^{-1}$ Mpc$^{-1}$.  To accelerate structure formation, we used an artificially enhanced normalization of the power spectrum, $\sigma_8=1.4$.  As discussed in \cite{stacyetal2010}, we verified that the density and velocity fields in the center of the minihalo closely resembled those in previous simulations.  We furthermore found that the angular momentum profile of the minihalo gas immediately before initial sink formation was very similar to the cosmological simulations of \cite{abeletal2002} and \cite{yoshidaetal2006}, despite their lower values of  $\sigma_8$ (0.7 and 0.9, respectively).  

The high resolution of our simulation was achieved through a standard hierarchical zoom-in procedure (see \citealt{stacyetal2010} for more details).  We added three additional nested refinement levels of 40, 30, and 20 kpc (comoving), centered on the location where the first minihalo will form.  At each level of refinement, we replaced every `parent' particle with eight `child' particles, such that at the highest refinement level each parent particle has been replaced with 512 child particles.  The highest-resolution gas particles have a mass $m_{\rm SPH} =0.015$~M$_{\odot}$, so that the mass resolution is  $M_{\rm res}\simeq 1.5 N_{\rm neigh} m_{\rm SPH} \la  1$ M$_{\odot}$, where $N_{\rm neigh}\simeq 32$ is the typical number of particles in the SPH smoothing kernel (e.g. \citealt{bate&burkert1997}).

\subsection{Cut-Out Technique}
Inclusion of feedback significantly reduces the simulation timesteps as compared to our control case with no feedback.  To facilitate faster computation speeds in our `with-feedback' case, once the main sink becomes massive enough to emit ionizing radiation we implement a `cut-out' technique.  In particular, we remove from the simulation box all particles that are located beyond 10 pc (physical) from the main sink, thereby following only the gravitationally bound central gas and discarding the very slowly evolving outer regions.  The cut-out region corresponds to the central $\la$ 4000 M$_{\odot}$ of gas. This technique reduces total computation time by nearly an order of magnitude while having a minimal effect on the simulated accretion history of the main sink, since by this point the central gas is dense, self-gravitating and no longer influenced by the gravity of the outer minihalo or DM on larger scales.  

We furthermore note that the gas at the edge of the 10 pc `cut-out' region has a typical density of $\sim 10^2$ cm$^{-3}$, corresponding to a free-fall time of $\sim 10^7$ yr, and thus should undergo little evolution within our simulation time of 5000 yr.  
In addition, though our method leads to a vacuum boundary condition at the edge of the cut-out, this should not be problematic.  The boundary conditions may lead to the propagation of a rarefaction wave starting from the cut-out edge, but this will only travel a distance of $c_{\rm s} \, t$, where $c_{\rm s}$ is the gas soundspeed ($\sim$ 2 km s$^{-1}$), and the time $t$ is 5000 yr.  This corresponds to a distance of $\sim$  10$^{-2}$ pc (2000 AU) from the cut-out edge, a very small distance compared to the 10 pc box size.  

\subsection{Chemistry, Heating, and Cooling}
We use the same chemistry and cooling network as described in \cite{greifetal2009}.   The code follows the abundance evolution of  H, H$^{+}$, H$^{-}$, H$_{2}$, H$_{2}^{+}$, He, He$^{+}$, He$^{++}$, and e$^{-}$, as well as the three deuterium species D, D$^{+}$, and HD. All relevant cooling mechanisms are accounted for, including H$_2$ cooling through collisions with He and H atoms and other H$_2$ molecules.  Also included are cooling through H and He collisional excitation and ionization, recombination, bremsstrahlung, and inverse Compton scattering.  We finally note that  H$_2$ cooling through collisions with protons and electrons is included, as they play an important role within H~{\sc ii} regions (\citealt{glover&abel2008}).    

One important difference we note between the high density ($n \ga 10^9$ cm$^{-3}$) evolution in the current simulation and that of \cite{stacyetal2010} is that we here account for the optical thickness of the H$_2$ ro-vibrational lines, which reduces the effectiveness of these lines in cooling the gas.  We include this effect using the Sobolev approximation (see \citealt{yoshidaetal2006, greifetal2011} for more details).   
For the three-body reactions

\begin{displaymath}
\rm H + H + H \rightarrow H_2 + H
\end{displaymath}
\noindent and
\begin{displaymath}
\rm H + H + H_2 \rightarrow H_2 + H_2
\end{displaymath}

\noindent we choose the rate coefficients adopted by \cite{pallaetal1983}. We note, however, that these reaction rates are still
subject to significant uncertainties (see \citealt{turketal2011} for a discussion).

\subsection{Sink Particle Method}
When an SPH particle reaches a density of $n_{\rm max} = 10^{12}$ cm$^{-3}$, we convert it to a sink particle.  If a gas particle is within the accretion radius $r_{\rm acc}$ of the sink, and if it is not rotationally supported against infall onto the sink, the sink accretes the particle.  The sink thus accretes the particles within its smoothing kernel immediately after it first forms.  We check for rotational support by comparing the angular momentum of the nearby gas particle,  $j_{\rm SPH} = v_{\rm rot} d$, with the angular momentum required for centrifugal support,  $j_{\rm cent} = \sqrt{G M_{\rm sink} r_{\rm acc}}$, where $v_{\rm rot}$ and $d$ are the
rotational velocity and distance of the particle relative to the sink. If a gas particle satisfies both  $d < r_{\rm acc}$ and $j_{\rm SPH} < j_{\rm cent}$, it is removed from the simulation, and the mass of the accreted particle is added to that of the sink.  

Our sink accretion algorithm furthermore allows for the merging of two sink particles.  We use similar criteria for sink particle merging as for sink accretion.  If the smaller, secondary sink is within $r_{\rm acc}$ of the more massive sink and has specific angular momentum less than $j_{\rm cent}$ of the larger sink, the sinks are merged.  After an accretion or merger event, the position of the sink is set to the mass-weighted average of the sink's former position and that of the accreted gas or secondary sink.  The same is done for the sink velocity.   We note that, as discussed in \cite{greifetal2011}, modifications to the sink merging algorithm can significantly alter the sink accretion history.  Future work will include studies of how a different technique for sink merging would modify our results.

We set the accretion radius equal to the resolution length of the simulation, $r_{\rm acc} = L_{\rm res} \simeq 50$ AU, where 

\begin{equation}
L_{\rm res}\simeq 0.5 \left(\frac{M_{\rm res}}{\rho_{\rm max}}\right)^{1/3} \mbox{\ ,}
\end{equation}

\noindent with $\rho_{\rm max}\simeq n_{\rm max}m_{\rm H}$ and $m_{\rm H}$
being the proton mass.  The sink particle's mass, $M_{\rm sink}$, is
initially close to the resolution mass of the simulation, $M_{\rm
res} \simeq 0.7$ M$_{\odot}$.
Sink particles are held at a constant density and temperature of  $n_{\rm max}$ = 10$^{12}$ cm$^{-3}$ and 650 K, such that their pressure is also kept to the corresponding value.  Providing the sink with a temperature and pressure prevents the existence of a pressure deficit around the sink, which would otherwise lead to artificially high accretion rates (e.g. \citealt{bateetal1995,brommetal2002,marteletal2006}).  The sink particles do continue to evolve in position and velocity, however, through gravitational and hydrodynamic interactions.

As discussed in \cite{brommetal2002} and \cite{stacyetal2010}, our sink formation criteria well represent regions that will truly collapse to stellar densities.  Before crossing the density threshold to become a sink, a gas particle must collapse two orders of magnitude above the average disk density, $\simeq 10^{10}$ cm$^{-3}$.  Our high density threshold and small value for  $r_{\rm acc}$ ensure that sinks are formed only from gravitationally collapsing gas.

Following the long-term evolution of the star-forming gas would not be feasible without the sink particle method.  By preventing gas evolution to ever higher densities, we avoid the problem of increasingly small numerical timesteps, a problem also known as `Courant myopia.'  We thus are able to see how the surrounding region of interest evolves over many dynamical times.  With sink particles, we can furthermore bypass the need to incorporate the chemistry, hydrodynamics and radiative transfer that comes into play at extremely high densities ($n>10^{12}$ cm$^{-3}$).  Finally, sink particles provide a convenient way to directly measure the accretion rate onto the protostellar region.

\begin{figure*}
\includegraphics[width=.8\textwidth]{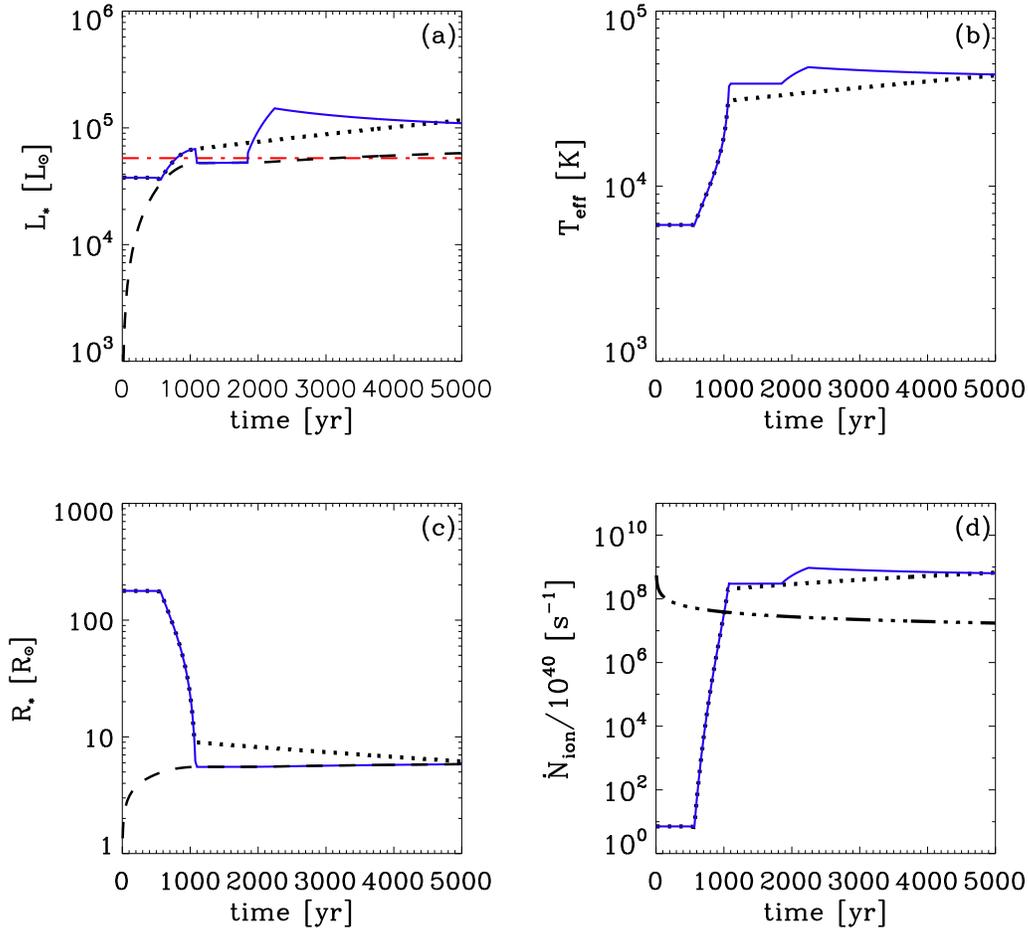}
 \caption{Evolution of various properties of the growing protostar according to our analytical model.  Solid blue lines represent the protostellar values used in our `with-feedback' simulation.  Dashed lines represent the ZAMS stellar values for the current sink mass.  Dotted lines represent the `slow contraction' case in which the accretion rate initially evolves in the same fashion as in the `with-feedback' simulation, but then holds steady at $10^{-3}$ M$_{\odot}$ yr$^{-1}$.
{\it (a):} Protostellar luminosity.  Red dash-dotted line is the estimate for $L_{\rm KH}$ of a 15 M$_{\odot}$ star of radius 10  R$_{\odot}$. 
{\it (b):} Effective temperature.
{\it (c):} Protostellar radius.
{\it (d):} Ionizing luminosity, $\dot{N}_{\rm ion}$.  Dashed-triple-dotted line represents the accretion rate of neutral particles onto the sink, extrapolated from the powerlaw fit to the sink accretion rate over the first 500 years 
($\dot{M} \propto t_{\rm acc}^{0.56}$, see section 3.5).    
Note how 
setting $R_* = R_{\rm ZAMS}$ after 1000 yr  
in our simulation yields a good approximation to the more physically realistic `slow-contraction' case.  Both also predict a break-out of ionizing radiation beyond the sink
at $\sim$ 1000 yr, when $\dot{N}_{\rm ion}$ 
exceeds the influx of neutral particles.
}
\label{star-model}
\end{figure*}


\subsection{Ray-tracing Scheme}
Once the first sink is formed, this particle is used as the source of protostellar LW and ionizing radiation. 
While the protostar is less massive than 10 M$_{\odot}$, LW radiation is the only source of feedback.  After the protostar is massive enough to 
emit ionizing radiation, however, a compact H~{\sc ii} region develops.  We model the growth of the surrounding I-front 
using a ray-tracing scheme which closely follows that of \cite{greifetal2009}.  A spherical grid with $\sim$ 10$^5$ rays and 200 radial bins is then created around the sink particle.   The minimum radius is determined by the distance between the sink and its closest neighboring SPH particle, and we update this structure each time the ray-tracing is performed. 
Because the sink accretes most particles within $r_{\rm acc} = 50$ AU, the minimum radius is usually $\ga 50$ AU.  The maximum radius is chosen as 
 10 pc (physical), the size of the cut-out simulation.  This value
easily encompasses the entire H~{\sc ii} region in our simulation.   The radial bins within 75 AU of the minimum radius are spaced at intervals of 1.5 AU.  Outside this distance the bins are logarithmically spaced.  The location of each particle is then mapped onto the corresponding bin within the spherical grid, and each particle contributes its density and chemical abundances to the bin proportional to its density squared.     

Next, the ionization front equation is solved along each ray:

\begin{equation}
n_{n}r_{\rm I}^{2}\frac{{\rm d}r_{\rm I}}{{\rm d}t}=\frac{\dot{N}_{\rm ion}}{4\pi}-\alpha_{\rm B}\int_{0}^{r_{\rm I}}n_{e}n_{+}r^{2}{\rm d}r\mbox{\ ,}
\end{equation} 

\noindent where $n_{n}$, $n_{e}$, and $n_{+}$ are the number densities of neutral particles, electrons, and positively charged ions, respectively.  The location of the ionization front with respect to the sink is denoted by $r_{\rm I}$.  $\dot{N}_{\rm ion}$ is the number of ionizing photons emitted per second, and $\alpha_{\rm B}$ is the case B recombination coefficient.  We use $\alpha_{\rm B} = 1.3 \times 10^{-12}$  cm$^{3}$ s$^{-1}$ for He~{\sc iii} recombinations to He~{\sc ii}, and $\alpha_{\rm B} = 2.6 \times 10^{-13} $  cm$^{3}$ s$^{-1}$ for He~{\sc ii} and H~{\sc ii} recombinations to the ground state (\citealt{osterbrock&ferland2006}).

The above I-front equation assumes that the ionization front is expanding into neutral, non-molecular gas.  This turns out to be an accurate assumption despite that the star is surrounded by a molecular disk.  As will be further described below, the ionization front expands only  into the lower-density polar regions which are indeed non-molecular.  Note that the gas does not become fully molecular until it reaches densities of $\sim 10^{10}$  cm$^{-3}$, and the average density of the ionized region is lower than this.  Furthermore, the ionization front is preceded by a photodissociation front whose extent always exceeds that of the ionized region, further ensuring that our ray-tracing scheme can safely ignore the presence of molecular species. 

The emission rate of H~{\sc i}, He~{\sc i} and He~{\sc ii} ionizing photons is given by

\begin{equation}
\dot{N}_{\rm ion}=\frac{\pi L_{*}}{\sigma_{\rm SB} T_{\rm eff}^{4}}\int_{\nu_{\rm min}}^{\infty}\frac{B_{\nu}}{h\nu}{\rm d}\nu\mbox{\ ,}
\end{equation}

\noindent where $h$ is Planck's constant, $\sigma_{\rm SB}$ the Stefan-Boltzmann constant, and $\nu_{\rm min}$ the minimum frequency required for ionization of the relevant species (H~{\sc i} or He~{\sc ii}).  For simplicity we do not distinguish between the H~{\sc ii} and He~{\sc ii} regions.  We assume the sink emits a blackbody spectrum $B_{\nu}$ with an effective temperature $T_{\rm eff}$, which depends upon the evolving stellar radius and luminosity. 

As described in \cite{greifetal2009}, the integral on the right-hand side of the ionization front equation is discretized by the following sum:

\begin{equation}
\int_{0}^{r_{\rm I}}n_{e}n_{+}r^{2}{\rm d}r\simeq\sum_{i}n_{e,i}n_{+,i}r_{i}^{2}\Delta r_{i}\mbox{\ ,}
\end{equation}

\noindent  where $\Delta r_{i}$ is the radial extent of each bin $i$, and the sum ranges from the sink particle to the current position of the I-front.  The left hand side of the ionization front equation is similarly discretized by: 

\begin{equation}
n_{n}r_{\rm I}^{2}\frac{{\rm d}r_{\rm I}}{{\rm d}t}\simeq\frac{1}{\Delta t}\sum_{i}n_{n,i}r_{i}^{2}\Delta r_{i} \mbox{\ ,}
\end{equation}

\noindent where the sum now extends from the I-front position at the previous timestep $t_0$ to its position at the current timestep  $t_0 + \Delta t$.  The above-described ray-tracing is performed separately for the H~{\sc ii} and He~{\sc iii} regions.

 This ray-tracing scheme utilizes the simplifying `on-the-spot' approximation (e.g. \citealt{osterbrock&ferland2006}), which we will argue in Section 3.3 is a sufficiently accurate assumption.  Our scheme also does not separately account for ionization by photons produced from recombination of He~{\sc ii} to  He~{\sc i}.  Instead, Equation 2 includes He~{\sc ii} in the $n_{+}$ term.  For simplicity, in terms of I-front evolution, He~{\sc i} and H~{\sc i} are effectively treated as the same species, as are He~{\sc ii} and H~{\sc ii}.  
  Nevertheless, the helium abundance is small, $\sim 0.08$, and the typical He~{\sc ii} abundance in the ionized region is slightly less than this value.  Ionizing photons from He~{\sc ii} recombination will thus be $\la$ 0.1 of those from H~{\sc ii} recombination, so the role of  He~{\sc ii} in the I-front evolution is relatively insignificant.  For our star, which typically has $T_{\rm eff} =$ 40,000 K (see Section 2.7), the He~{\sc ii} and H~{\sc ii} regions indeed have the same radial extent as expected (e.g. \citealt{osterbrock&ferland2006}), while $T_{\rm eff}$ is too low for an He~{\sc iii} region to form.

 We note that we do not resolve higher-density regions within the sink, and thus do not directly simulate the disk self-shielding and absorption of ionizing photons on this scale (e.g. \citealt{mckee&tan2008}).  In order to model the effect of self-shielding on sub-sink scales, we set the I-front radius to zero along all rays that encounter bins with density greater than $5 \times 10^9$ cm$^{-3}$.   This is the typical density of gas within the disk (see Section 3.1), and our prescription thus allows the directions along the dense molecular disk to be shielded while the more diffuse polar regions are the first to become ionized.  This simple modeling of self-shielding also assumes that the large-scale disk surrounding the sink has the same orientation as the unresolved sub-sink disk.  

\subsection{Photoionization and Heating}
Particles determined to be within the extent of the  H~{\sc ii} and  He~{\sc iii} regions are given additional ionization and heating rates in the chemistry solver:

\begin{equation}
k_{\rm ion}=\int_{\nu_{\rm min}}^{\infty}\frac{F_{\nu}\sigma_{\nu}}{h\nu}{\rm d}\nu
\end{equation}

\noindent and

\begin{equation}
\Gamma=n_{n}\int_{\nu_{\rm min}}^{\infty}F_{\nu}\sigma_{\nu}\left(1-\frac{\nu_{\rm min}}{\nu}\right){\rm d}\nu\mbox{\ ,}
\end{equation}

\noindent where $F_{\nu}$ is the incoming specific flux and $\sigma_{\nu}$ the photo-ionization cross section. For a blackbody, we have

\begin{equation}
F_{\nu}=\frac{L_{*}}{4\sigma_{\rm SB} T_{\rm eff}^{4}r^{2}}B_{\nu}\mbox{\ ,}
\end{equation}

\noindent where $r$ is the distance from the sink.

Finally, H$_2$ dissociation by LW radiation (11.2 to 13.6 eV) is described by

\begin{equation}
k_{{\rm H}_{2}} =  1.1\times 10^{8}\,f_{\rm shield}\,F_{\rm LW}~{\rm s}^{-1} 
\end{equation}

\noindent  (\citealt{abeletal1997}), where $F_{\rm LW}$ denotes the radiation flux, in units of   
 erg s$^{-1}$ cm$^{-2}$ Hz$^{-1}$, at $h\overline{\nu}=12.87$eV,
and  $f_{\rm shield}$ is the factor by which  H$_{2}$ self-shielding reduces the LW dissociation rate.  This self-shielding factor depends upon the  H$_{2}$ column density $N_{\rm H_2}$.  With the above ray-tracing scheme, we determine $N_{\rm H_2}$ along each ray by summing the contribution from each bin.  We then use results from \cite{draine&bertoldi1996} to determine the value for $f_{\rm shield}$.  We note a recent update to their $f_{\rm shield}$ fitting formula (\citealt{wolcottetal2011,wolcott&haiman2011}), but do not expect this to significantly affect the results for our particular case.  Because of the unusually high column densities within the molecular disk ($N_{\rm H_2} \ga 10^{26}$ cm$^{-2}$), we calculate  $f_{\rm shield}$ using equation 37 from  \cite{draine&bertoldi1996}, which is more accurate  for large $N_{\rm H_2}$ than their simple power-law expression in their equation 36.    

These heating, ionization, and dissociation rates are accounted for at every timestep.  They are updated every 10 yr as they evolve with the protostellar mass and accretion rate, because these determine the values of $L_*$ and $T_{\rm eff}$.
The ray-tracing procedure is performed every fifth timestep.
 As discussed in \cite{whalen&norman2006}, it is important to update the ray-tracer often enough to correctly model the propagation speed of the I-front.  We discuss a test later in Section  3.3 to show that our ray-tracing procedure is performed with sufficient frequency to correctly model the I-front growth.
 
Our numerical method also neglects heating and ionization of gas ahead of the I-front by hard UV photons.  Thus, instead of a gradual increase in temperature, ionized gas is instantly  heated to $> 20,000$ K.  As discussed in \cite{whalen&norman2008} and (2008b)\nocite{whalen&norman2008b}, this may artificially suppress I-front instabilities.  This includes thin-shell instabilities driven by H$_2$ cooling.  These may be somewhat mitigated by LW radiation, however, particularly at the early times studied in our simulation when the I-front remains in relatively close proximity to the star ($\la 0.1$ pc) and H$_2$ shielding at the I-front edge is not too severe ($N_{\rm H_2} \sim 10^{16}$ cm$^{-2}$).  One may guess that such instabilities would allow ionizing radiation to be more easily channeled away from the stellar disk, allowing greater mass inflow onto the star (e.g. Whalen \& Norman 2008b).  Though currently too computationally expensive, a future study of I-front instabilities on sub-parsec scales using multifrequency radiative transfer would be highly worthwhile.

\subsection{Radiation Pressure}

For simplicity, we do not include effects of radiation pressure in our calculation.  We can estimate the effect of radiation pressure due to Thomson scattering by comparing the typical protostellar luminosity ($L_* =10^5$ L$_{\odot}$, Fig. \ref{star-model}) with the Eddington luminosity: $L_{\rm Edd} = 4\pi G M_* m_{\rm p} c/\sigma_{\rm T}$, where $M_*$ is the stellar mass, $m_{\rm p}$ is the mass of a proton, and $\sigma_{\rm T}$ is the Thomson scattering cross section.  For our typical stellar mass of 20 M$_{\odot}$, $L_{\rm Edd} \simeq 6 \times 10^5$ L$_{\odot}$.  $L_*$ is several times smaller than this, so we would not expect Thomson scattering to be dynamically important.  

Following similar discussion in, e.g., \cite{haehnelt1995,mckee&tan2008}, and given balance between ionizations and recombinations, we estimate the distance from the protostar at which pressure from ionizing radiation will dominate over gravity by comparing radiative and effective gravitational forces:

\begin{eqnarray}
F_{\rm ion} &=& F_{\rm grav,eff} \nonumber\\
\alpha_{\rm B} n_{\rm p} \left(\frac{h\nu_i}{c}\right)   &=& \frac{\phi_{\rm Edd}GM_*\mu m_{\rm H} }{r^2}
 \end{eqnarray}
 
 \noindent where $\phi_{\rm Edd} = 1 - L_*/L_{\rm Edd} \simeq 0.83$.  Using typical H~{\sc ii}  region densities of 10$^{7}$ cm$^{-3}$ and $h\nu_i = 13.6$ eV, $F_{\rm ion}$ will become greater than $F_{\rm grav,eff}$ beyond distances of $\sim$ 100 AU, similar to the scale of the gravitational radius $r_{\rm g}$.  However, beyond  $r_{\rm g}$ gas pressure should begin to dominate over both gravity and radiation pressure.  Gas pressure is approximately given by $nk_{\rm B}T \ga 3 \times 10^{-5}$  dyn cm$^{-2}$ within the 20,000 K H~{\sc ii}  region.  Ionization pressure is approximately 
 
 \begin{equation}
 P_{\rm ion} \sim  \frac{\dot{N}}{4\pi r^2} \left(\frac{h\nu_i}{c}\right) 
 \end{equation}
 
 \noindent which is $\sim 3 \times 10^{-5}$  dyn cm$^{-2}$ for $r = 200$ AU and $\dot{N} = 5 \times 10^{48}$ s$^{-1}$.  $P_{\rm ion}$ will rapidly fall with radius, so on the scales which we study of 50 AU and beyond, gas pressure will usually dominate over ionizing radiation pressure.  However, future work should include the effects of $P_{\rm ion}$,  particularly studies which examine very early I-front growth on scales below $\sim$ 100 AU.  

 Let us also estimate another potentially important effect, radiation pressure from Lyman-$\alpha$ (Ly$\alpha$) photons.  Assuming opaque conditions, pressure from Ly$\alpha$ can be written as 
$P_\alpha=\frac{1}{3} u_\alpha= 4\pi J_\alpha/3c$
where $u_\alpha$ and $J_\alpha$ are the energy density and mean intensity of
the Ly$\alpha$ radiation (see discussion in \citealt{mckee&tan2008}).  A rough estimate is found in, e.g., \cite{bithell1990, oh&haiman2002},  

\begin{equation}
P_\alpha = \frac{\dot{N}_{\rm Ly\alpha}}{4\pi r^2} \left(\frac{h \nu}{c} \right) \frac{s_{\tau}}{\tau_{\rm Ly\alpha}}
\end{equation}

\noindent where $\dot{N}_{\rm Ly\alpha}$ is the rate of emission of Ly$\alpha$ photons by the star and $s_{\tau}$ is
 the Ly$\alpha$ photon path length in units of the optical depth of the region of gas in question, $\tau_{\rm Ly\alpha}$ (see also discussion and more detailed equation in \citealt{mckee&tan2008}).  From equation 21 of \cite{bithell1990}, we estimate  $s_{\tau}/ \tau_{\rm Ly\alpha}$ to be $\sim$ 5000 for the disk gas and $\simeq$ 100 for the ionized region.  However, this may be further reduced by the motion of the gas (e.g. \citealt{bithell1990,haehnelt1995,mckee&tan2008}).  The Ly$\alpha$ photons thus travel much more freely through the ionized region.  Estimating $\dot{N}_{\rm Ly\alpha} \sim \dot{N} \sim 5 \times 10^{48}$ s$^{-1}$ for the $\sim$ 20  M$_{\odot}$ protostar, we then find upper limits of $P_\alpha \sim 5 \times 10^{-3}$  dyn cm$^{-2}$at the edge of the 1000 AU disk, and $10^{-5}$ dyn cm$^{-2}$ at the edge of the 10$^4$ AU H~{\sc ii} region.

As discussed in detail in \cite{mckee&tan2008}, Ly$\alpha$ radiation pressure can be estimated to significantly slow protostellar accretion when it  is over twice the ram pressure of infalling gas, $P_{\alpha} > 2\rho v_{\rm ff}^2$, where $v_{\rm ff}$ is the free-fall velocity of infalling gas.  For gas in a rotating disk, the requirement becomes $P_{\alpha} > \rho v_{\rm Kep}^2 = \rho v_{\rm ff}^2/2$. At the disk edge where $n \sim 10^9$ cm$^{-3}$, $\rho v_{\rm Kep}^2 \sim 10^{-3}$ dyn cm$^{-2}$.  At the edge of the H~{\sc ii} region where $n \sim 10^7$ cm$^{-3}$, we have $2\rho v_{\rm ff}^2 \la 10^{-5}$ dyn cm$^{-2}$.  These values are indeed exceeded by the corresponding strength of $P_\alpha$.  In contrast, within a few hundred AU of the main sink where  $n \sim 10^{10} - 10^{11}$ cm$^{-3}$,  $P_\alpha$ is up to several times smaller than $\rho v_{\rm Kep}^2$, and the gravitational force of the sink is more dominant.  

Consistent with the findings of \cite{mckee&tan2008}, Ly$\alpha$ may indeed break out when a star has reached 20 M$_{\odot}$, or upon reaching even larger masses for the case of  gas undergoing less rotation.   Ly$\alpha$ pressure is thus expected to break out only after the star  has become massive enough to develop an  H~{\sc ii} region.  We would furthermore expect Ly$\alpha$ to break out in the polar regions before affecting the disk plane, as the polar regions are more diffuse.  Afterwards, Ly$\alpha$ will easily escape through the polar cavities, relieving pressure closer to the disk plane through which the protostars accrete.  We thus do not expect that inclusion of Ly$\alpha$ pressure would significantly change our results, though future work should confirm this through more detailed simulations.


\begin{figure*}\includegraphics[width=.8\textwidth]{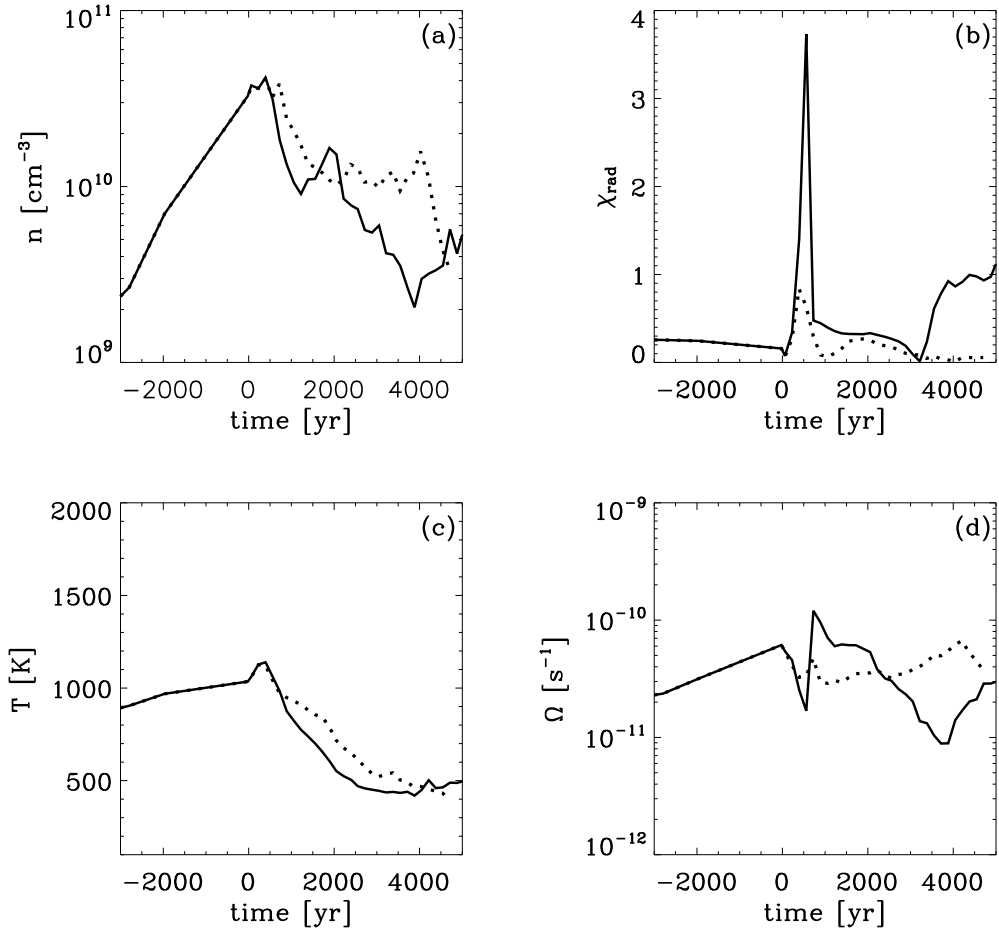}
\caption{Evolution of various disk properties.  Time is measured from the point at which the first sink particle forms.  In each case, solid lines represent the `no-feedback' case while the dotted lines denote the `with-feedback' case.
{\it (a):} The average density of the gas within the disk.  
{\it (b):} The ratio of the average radial to the average rotational velocity of the disk particles, $\chi_{\rm rad}$.  
{\it (c):} The average temperature of the disk particles.  
{\it (d):} The average angular velocity of the disk gas with respect to the disk center of mass, $\Omega$.  Note how the `with-feedback' and `no-feedback' cases diverge after 1000 yr.}
\label{disk}
\end{figure*}

\subsection{Protostellar evolution model}

Our ray-tracing algorithm requires an input of protostellar effective temperature $T_{\rm eff}$ and luminosity $L_*$.  This is the sum of $L_{\rm acc}$, the accretion luminosity,
and $L_{\rm photo}$, the luminosity generated at the protostellar surface:

\begin{eqnarray}
L_* = L_{\rm acc} + L_{\rm photo} =  \frac{G M_* \dot{ M}}{R_*} + L_{\rm photo}\mbox{,}
\end{eqnarray}

\noindent where $M_{*}$ is the protostellar mass, $\dot{M}$ the accretion rate,  and $R_*$ the protostellar radius.
The photospheric luminosity is either due to Kelvin Helmholtz (KH) contraction, or due to hydrogen burning which begins once the protostar has reached the zero-age main sequence (ZAMS). 
For simplicity, we set $L_{\rm photo}\simeq L_{\rm ZAMS}$,  thus assuming a robust upper limit in the luminosity at later phases in the protostellar evolution.  Initially, however, we assume no contribution from $L_{\rm photo}$ until $R_*$ reaches the ZAMS radius. This is a reasonable assumption, particularly given the values of $L_{\rm photo}$ as determined by, e.g., \cite{hosokawaetal2010}.  We find that  $L_{\rm acc}$ should be much greater or similar in magnitude to $L_{\rm photo}$ until the ZAMS is reached in our model (see Fig. \ref{star-model}). Our assumption is also robust in view of uncertainties regarding how rapidly KH contraction proceeds.  While \cite{hosokawaetal2010} find that a massive ($\sim$ 10 M$_{\odot}$) accreting protostar will typically contract to the ZAMS radius on timescales of $\sim 10^{4}$ yr, in our model we set our protostar to the ZAMS values much earlier.  However, the luminosity during KH contraction is indeed similar in magnitude to $L_{\rm ZAMS}$ for a star of the same mass (\citealt{hosokawaetal2010}).  
Nevertheless, it is important that future simulations self-consistently couple protostellar evolution and accretion flow under feedback.

As in \cite{schalleretal1992} and \cite{hosokawaetal2010}, we determine $L_{\rm ZAMS}$ using a simple fit to the stellar mass:

\begin{equation}
L_{\rm ZAMS} = 1.4 \times 10^4 {\rm L_{\odot}} \left(\frac{M_*}{10 \rm M_{\odot}}\right)^2 \mbox{\ .}
\end{equation}

\noindent Each time $L_*$ is updated, we assume $M_{*}$ is equal to the sink mass.  Due to the discrete nature of sink accretion in an SPH simulation, instead of calculating $\dot{M}$ at each timestep, we determine $\dot{M}$ by averaging the sink mass growth over the previous 100 yr, updating $\dot{M}$ every 10 yr.  

We estimate $R_*$ in the same fashion as in \cite{stacyetal2010}, which was based on the prescription of \cite{omukai&palla2003}.  We find that during the adiabatic accretion phase, $R_*$ grows as

\begin{equation}
R_{*I} \simeq 50 {\rm R_{\odot}} \left(\frac{M_*}{\rm M_{\odot}}\right)^{1/3} \left(\frac{\dot{M}}{\dot{M}_{\rm fid}}\right)^{1/3}   \mbox{\ ,}
\end{equation}
      
\noindent where $\dot{M}_{\rmn fid}\simeq 4.4\times 10^{-3} {\rmn M}_{\odot}$\,yr$^{-1}$ is a fiducial rate, typical for Pop~III accretion.  Throughout this phase, 
we assume there is not yet any contribution from  $L_{\rm photo}$.
During the subsequent phase of KH contraction, the radius will shrink according to

\begin{equation}
R_{*II} \simeq 140 {\rmn R_{\odot}} \left(\frac{\dot{M}}{\dot{M}_{\rmn fid}}\right) \left(\frac{M_*}{10 \rmn M_{\odot}}\right)^{-2} \mbox{\ .}
\end{equation}

\noindent We estimate that the transition from adiabatic accretion to KH
contraction occurs when the value of $R_{*II}$ falls below that of
$R_{*I}$.  During this phase, 
our model again assumes no luminosity contribution from  $L_{\rm photo}$, and that $L_{\rm acc}$ is the main contribution to the luminosity.  
KH contraction will halt once the star has reached the ZAMS,
at which point we set $R_*$ equal to the ZAMS radius,

\begin{equation}
R_{\rm ZAMS} =   3.9 {\rm R_{\odot}}  \left(\frac{M_*}{10 \rm M_{\odot}}\right)^{0.55}
\end{equation}

\noindent (e.g. \citealt{hosokawaetal2010}).  We set $R_*$ equal to $R_{\rm ZAMS}$ when the value for $R_{*II}$ falls below $R_{\rm ZAMS}$.

If the calculated accretion rate drops to near zero, then the radial values for the adiabatic and KH contraction phases will become vanishingly small.  If this occurs before the sink has been accreting for a KH time and reached the ZAMS, the accretion rate is set to the previous non-zero value in order to get more realistic values for $R_{*I}$ and $R_{*II}$.   This allows us to avoid setting $R_* = R_{\rm ZAMS}$ too early in the protostar's evolution.  If, however, the accretion slows after the sink has been in place for more than its KH time, we assume the star has reached its ZAMS radius, and we set $L_{\rm acc} = 0$, $R_* = R_{\rm ZAMS}$, and $L_* = L_{\rm ZAMS}$.  Note that typical KH times, where $t_{\rm KH} = G M_*^2/ R_* L_*$, range from 1000 yr for a large and rapidly accreting 10 M$_{\odot}$ protostar (see e.g. \citealt{hosokawaetal2010}) to $\sim 4 \times 10^4$ yr for a 15 M$_{\odot}$ main sequence star. The typical KH luminosity for a 15 M$_{\odot}$ star is $L_{\rm KH} \sim 5 \times 10^{4}$ L$_{\odot}$ (see Fig. \ref{star-model}).    

\begin{figure*}
\includegraphics[width=.45\textwidth]{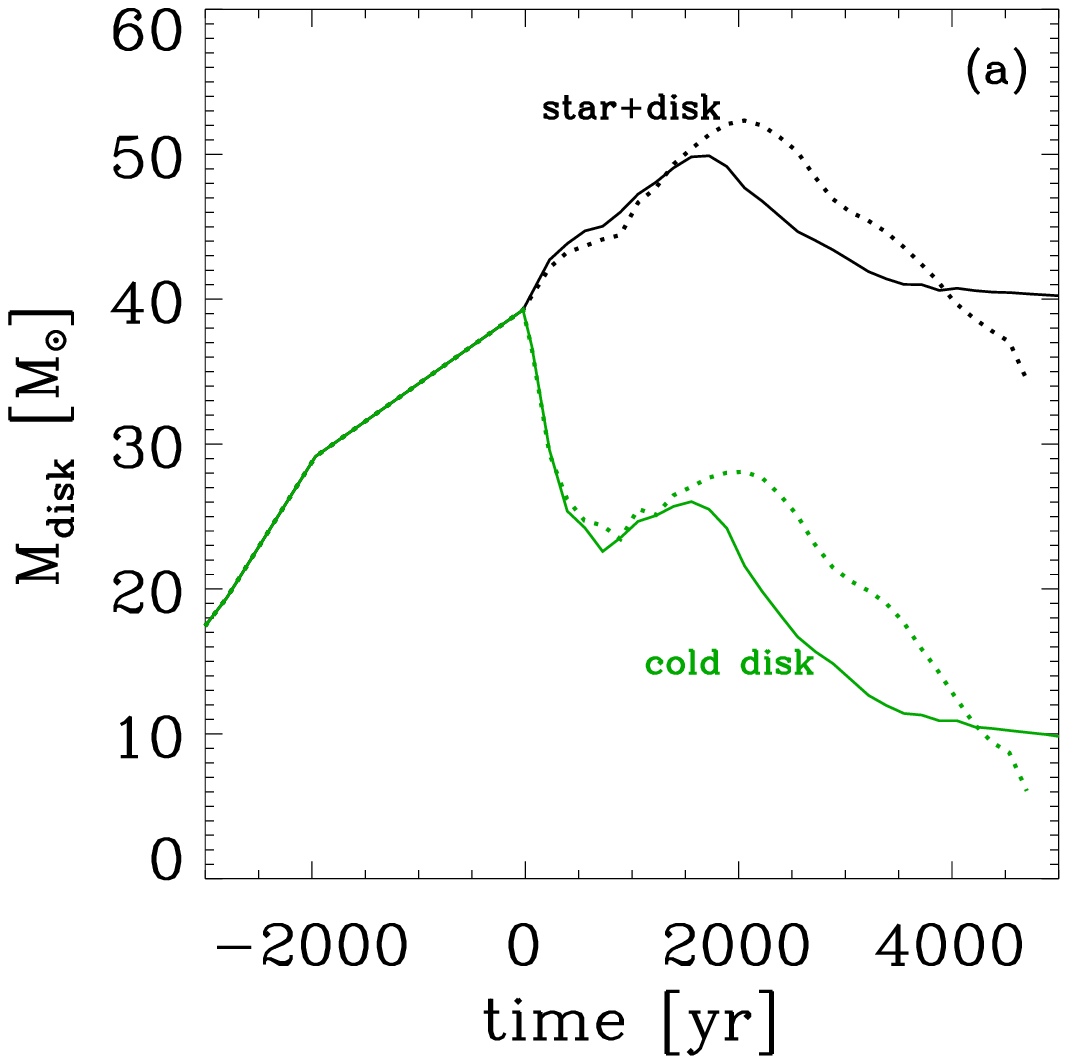}
\includegraphics[width=.45\textwidth]{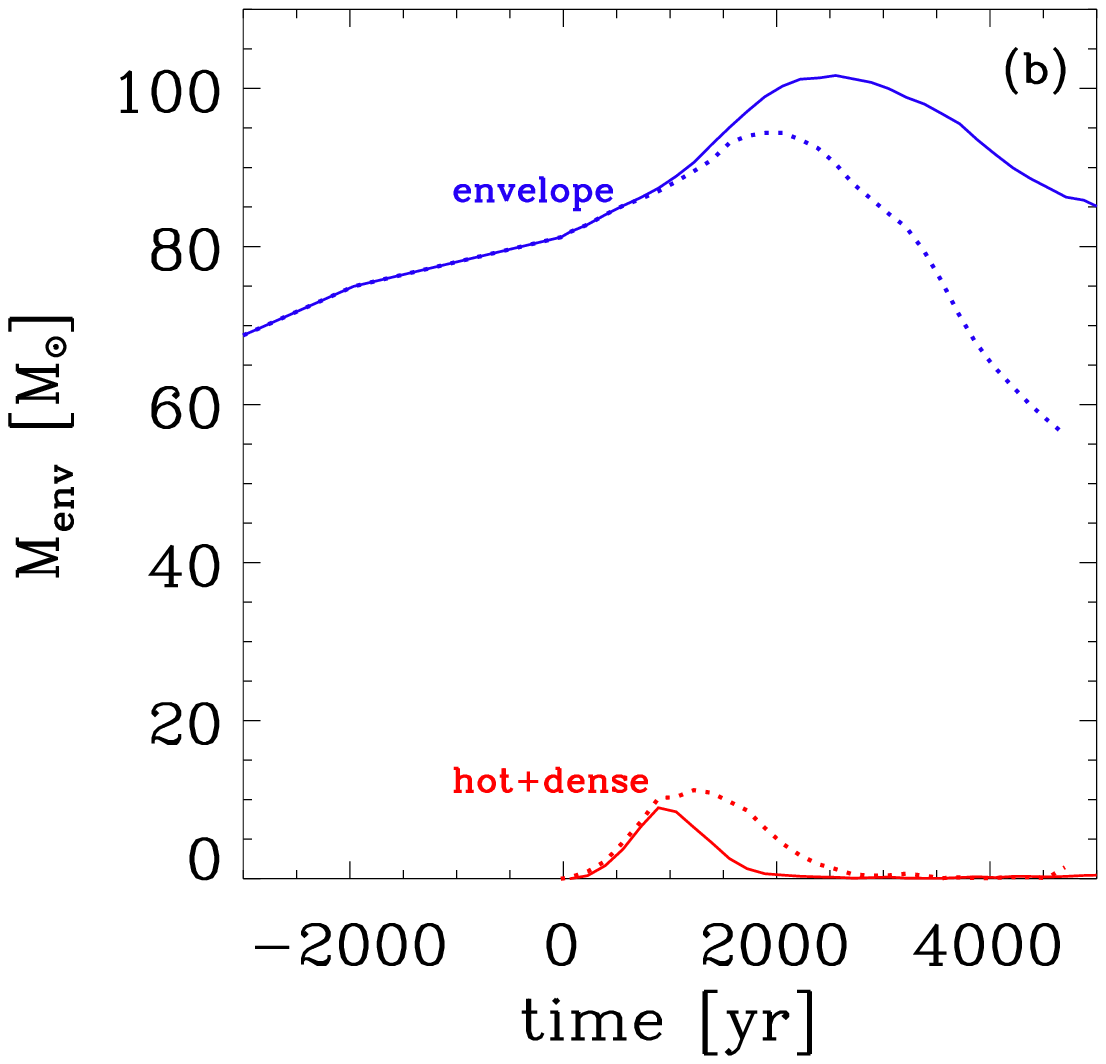}
\caption{Evolution of disk mass over time.  Solid lines are for the `no-feedback' case.  Dotted lines represent the `with-feedback' case. 
{\it (a):} Black lines show the mass of the total star-disk system.  Disk mass (green lines) is taken as dense ($n > 10^9$ cm $^{-3}$) gas with an H$_2$ fraction greater than 10$^{-3}$.  
{\it (b):} Red lines show the mass of hotter, non-molecular material in the same density range as the disk, and blue lines denote the total mass of the outer envelope, which is defined as comprising all sinks and $n > 10^8$ cm $^{-3}$ gas.  Note how radiative feedback 
 causes the total disk and envelope mass to decline after approximately 2000 yr.  N-body dynamics in the `no-feedback' case causes an even steeper initial decline in disk mass compared to radiation in the `with-feedback' case.}
\label{diskmass}
\end{figure*}

Given our averaging scheme in which a minimum of one 0.015 M$_{\odot}$ gas particle can be accreted over 100 years, this gives an effective minimum measurable accretion rate of $1.5 \times 10^{-4}$  M$_{\odot}$ yr$^{-1}$.  However, for $M_* \ga 10$~M$_{\odot}$, this minimum accretion rate still yields a value of  $R_{*II}$ that is smaller than  $R_{\rm ZAMS}$.  In this case, we again set the protostellar luminosity and radius to its ZAMS values once the accretion rate has dropped to 
 $\la$ $10^{-4}$  M$_{\odot}$ yr$^{-1}$.  
In our case, the measured accretion rate drops very quickly after 500 yr.  At this point the star has reached 15 M$_{\odot}$, is still undergoing adiabatic expansion, and has $t_{\rm KH} \sim 1000$ yr.  The star then begins rapid KH contraction until the measured accretion rate 
drops below $10^{-4}$  M$_{\odot}$ yr$^{-1}$ at $\ga$ 1000 yr, not rising above this value again until 
$\sim 2000$ yr (see increase in luminosity at this time in Fig. \ref{star-model}).  
Though within the simulation we set  $R_* = R_{\rm ZAMS}$ as soon as the averaged accretion rate 
falls below $10^{-4}$  M$_{\odot}$ yr$^{-1}$, 
in reality the protostar is better described by a more gradual approach to  $R_{\rm ZAMS}$.   In Figure \ref{star-model} we show the protostellar values used in the simulation along with a more realistic `slow-contraction' model which follows the same accretion history as the `with-feedback' case until reaching an asymptotic growth rate of $10^{-3}$  M$_{\odot}$ yr$^{-1}$.  The `slow-contraction' model is then held at this rate, which is similar to the fiducial value used in Equation 11, and is also the typical accretion rate found at late times in our `no-feedback case' as well as the simulations of, e.g., \cite{greifetal2011,smithetal2011}.  This model well-matches the prescription used in the simulation, particularly in effective temperature and ionizing luminosity, and both predict that ionizing radiation will exceed the influx of neutral particles, and that break-out beyond the sink occurs at $\sim$ 1000 yr.  Though the protostar most likely does not reach the ZAMS within the time of our simulation, our simple model serves as a reasonable approximation for any unresolved accretion luminosity.  

We also note that \cite{hosokawaetal2010} find that, for a given accretion rate, primordial protostars undergoing disk accretion will have considerably smaller radii than those accreting mass in a spherical geometry.  The rapid contraction of the protostar between 500 and 1000 years therefore serves as an idealized representation of the sub-sink material evolving from a spherical to a disk geometry as it gains angular momentum.          
We emphasize, however, that the unrealistically rapid contraction to the ZAMS in our model, along with the incomplete inner-disk self-shielding, leads to an overestimate in the strength of feedback.  In this way we underestimate the mass of the star.   
In addition, the recent work by \nocite{smithetal2011b} Smith et al. (2011b) found that Pop III protostars undergoing variable accretion rates may have very large radii (100-200 R$_{\odot}$) for most of their pre-main sequence lifetimes.  This again indicates a probable underestimate of the protostellar radius and an overestimate of $T_{\rm eff}$ in our model.  
However, note that Smith et al. (2011b) assumed a `hot' spherical accretion model (e.g. \citealt{hosokawaetal2010}, \nocite{hosokawaetal2011b} 2011b), which will yield significantly larger radii than the `cold' thin disk accretion model.  Though disk thickness as well as high disk temperatures and accretion rates may render `hot' accretion the more realistic case, the true radial evolution of the star is likely to be in between these two possibilities.  

As discussed in \cite{smithetal2011}, the sink particle method requires several simplifying assumptions when constructing the protostellar model.  
By setting the protostellar mass equal to the sink mass, we are assuming that the small-scale disk which likely exists within the sink region has low mass compared to the protostar.
We also assume that the accretion rate at the sink edge is the same as the accretion rate onto the star, when in reality after gas enters within $r_{\rm acc}$ it must likely be processed through a small-scale disk before being incorporated onto the star.  However, our assumption may still be a good approximation of physical reality, given that the primordial protostellar disk study of Clark et al. (2011b) \nocite{clarketal2011b} implies that the thin disk approximation would probably not be valid on sub-sink scales (e.g. \citealt{pringle1981}), and that strong gravitational torques can quickly drive mass onto the star.  Furthermore, as also pointed out in \cite{smithetal2011}, our averaging of the accretion rate over a number of timesteps serves to mimic the buffering of accreted material by the sub-sink disk.  The inputs to our protostellar model are thus necessarily approximate, as is the protostellar model itself.  
Nevertheless, it is sufficient to provide an exploratory picture of how ionizing feedback will affect Pop III accretion within a cosmological setup.

\section{Results}

\subsection{Disk Evolution}

The gas in both the `no-feedback' and `with-feedback' cases underwent disk formation, fragmentation, and the emergence of several sinks.   We show the evolution of various disk properties in Figure \ref{disk}.  Because of the imprecision involved in determining which gas particles comprise the disk, for simplicity we define the disk as consisting of particles with number density greater than 10$^9$ cm$^{-3}$ and with an H$_2$ fraction greater than 10$^{-3}$.  This way the disk only contains cool and dense gas that has not been subject to ionization or significant H$_2$ destruction.  From Figure \ref{diskmass} we see that the disk structure is growing in mass well before the first sink forms.  This central gas is already rotationally dominated, as indicated in Figure \ref{disk} by the low values of $\chi_{\rm rad}$, which is the average radial velocity of the gas particles divided by their average rotational velocity, $\chi_{\rm rad} = v_{\rm rad}/v_{\rm rot}$.  Velocities are measured with respect to the center of mass of the disk.

\subsubsection{No-feedback case}
After sink formation, the `no-feedback' disk steadily grows in mass (Fig. \ref{diskmass}) as angular momentum causes it to expand in radius and become somewhat lower in average density (panel {\it a} in Fig. \ref{disk}). The disk growth is halted at nearly 2000 yr due to its gradual disruption through N-body dynamics of the sinks.  One of the sinks is ejected at $\sim$ 500 yr, after growing to only $\sim$ 1 M$_{\odot}$.  The ejection occurs immediately following the merger of the two other sinks, at a time when the three sinks are close together in the center of the disk and subject to N-body dynamics. The sink accretes no more mass after its ejection.  It initially moves in a direction perpendicular to the disk plane at $\sim$ 5 km s$^{-1}$ with respect to the disk center of mass, and the maximum distance between the two sinks approaches 3000 AU at approximately 2500 yr. 
This increase in distance between the two sinks is mostly due to the motion of the main sink, however.  In this N-body interaction, the main sink gains a larger velocity, initially moving with respect to the disk center of mass at $\sim$ 10 km s$^{-1}$.  It travels parallel to the disk plane and pulls the disk along with it.  
The rapid motion of the main sink disrupts the high-density gas, which transforms from a disk structure to a more diffuse tidal tail, eventually causing the total measured disk mass to slightly decrease (Fig. \ref{diskmass}).  The rotational structure is also disturbed, as indicated by a peak in $\chi_{\rm rad}$ when the sink is ejected.   
After approximately 3000 yr, the tidal tail begins to recompress, causing the dense gas to be more dominated by radially inward motion.  This is indicated by the increase of $\chi_{\rm rad}$ to $\sim 1$ as well as an increase in the average disk density (Fig. \ref{disk}).   

\begin{figure*}
\includegraphics[width=.8\textwidth]{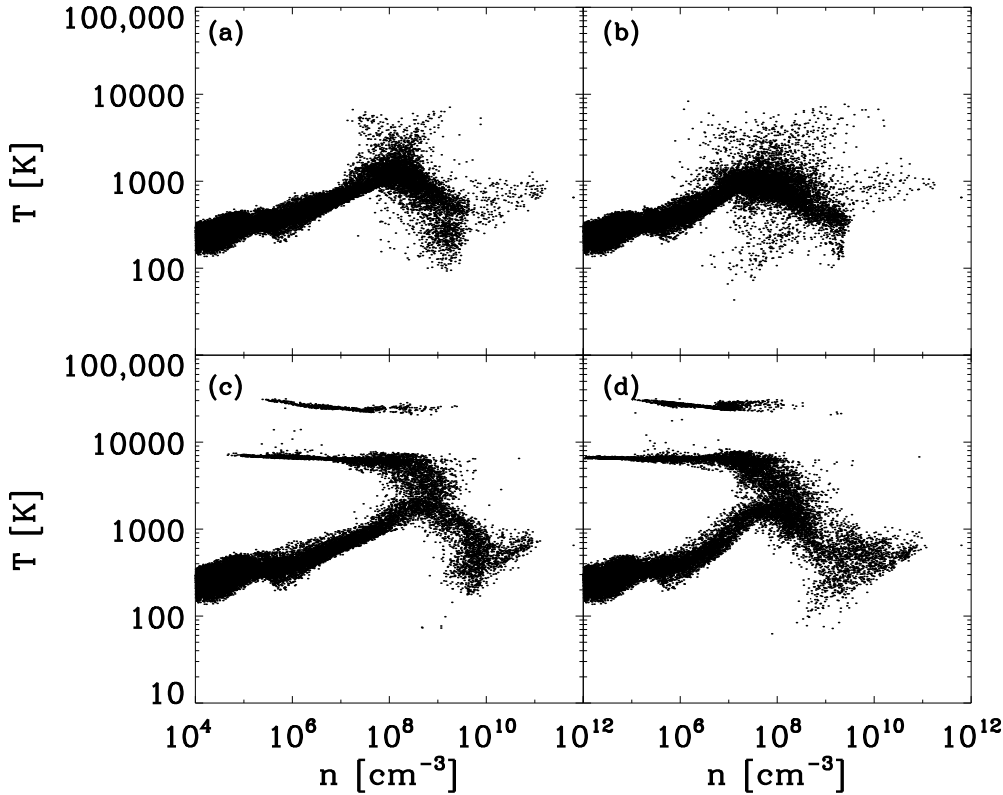}
\caption{Temperature versus number density for both cases at various times in the simulations.
{\it (a):} `No-feedback' case at 2500 yr.
{\it (b):} `No-feedback' case at 5000 yr.
{\it (c):} `With-feedback' case at 2000 yr.
{\it (d):} `With-feedback' case at 3000 yr.
Note how, in the `no-feedback' case, there is only a light stream of particles with $n > 10^9$ cm$^{-3}$ that is accreting onto the main sink. The gravitational potential well of the main sink leads to the heating of a growing region of lower-density gas.   In the `with-feedback' case, there is an expanding ionized region with temperature of 20,000 K along with a larger region of hot neutral gas with temperature of several thousand Kelvin.}
\label{Tvsnh}
\end{figure*}

From the solid red line in Figure \ref{diskmass}, we also note a smooth early growth of hot dense gas over the first 1000 yr.  This begins almost as soon as the first sink is formed, and the main sink thus provides an early source of heating. This occurs as the gravitational potential of the sink, which grows to $> 10$ M$_{\odot}$  in only 200 yr, heats the surrounding gas up to $\sim$ 7000 K, or $c_{\rm s} \simeq 7$ km s$^{-1}$ (Fig. \ref{Tvsnh}), corresponding to approximately the virial temperature of the sink:   
\begin{equation}
T_{\rm vir}\simeq \frac{G M_{\rm sink}m_{\rm H}}{k_{\rm B}r_{\rm acc}}
\simeq 10^4 \mbox{\,K.}
\end{equation}

\begin{figure*}
\includegraphics[width=.9\textwidth]{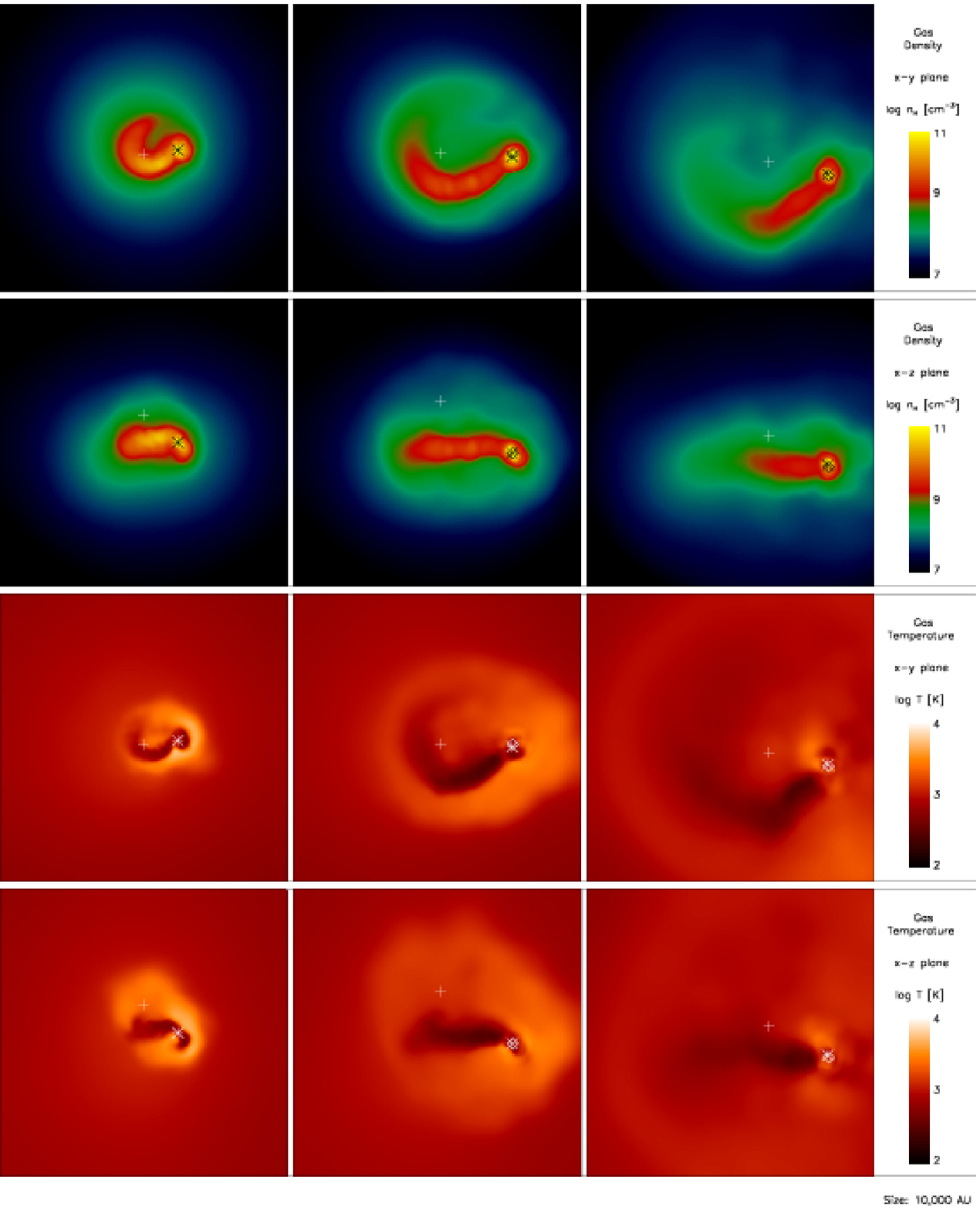}
\caption{Projected density and temperature structure of central 10,000 AU without protostellar feedback at 1000 yr ({\it left column}), 2000 yr ({\it middle column}), and 5000 yr ({\it right column}) after initial sink formation.  Asterisks denote the location of the most massive sink.  
Crosses show the location of the ejected sink. Diamonds mark the locations of the other sinks.  
$\it{Top}$ $\it{ row:}$ Density structure of the central region in the x-y plane. 
$\it{Second}$ $\it{row:}$ Density structure of the central region in the x-z plane.  
$\it{Third}$ $\it{row:}$ Temperature structure of the central region in the x-y plane.  
$\it{Bottom}$ $\it{row:}$ Temperature structure of the central region in the x-z plane.
Note the formation of a cool disk and tidal tail structure along with the growth of a surrounding bubble of warm gas. The ejected sink (cross) has a mass of 1~M$_{\odot}.$}
\label{morph_nf}
\end{figure*}

The mass of hot dense gas undergoes an equally smooth decline over the next 1000 yr in the `no-feedback' case.  A minimal amount of dense gas is newly heated after 1000 yr because there is no longer a dense disk structure entirely surrounding the main sink.  Most of the dense gas instead trails behind the sink in the tidal tail, and the gravitational heating provided by the sink is now imparted directly to the lower-density particles.  Sink gravitational heating thus halts the growth of the outer envelope, which we define to include all gas particles at densities above 10$^8$ cm$^{-3}$ as well as the sinks (solid blue line in Fig. \ref{diskmass}). The `heat wave' travels beyond the disk at approximately the sound speed, causing the decline in envelope mass beginning at around 2500 yr in Figure \ref{diskmass}, and reaching distances of $> 7000$ AU (see Figs. \ref{Tvsnh} and \ref{morph_nf}) by the end of the simulation.  At this time, heated gas thus resides almost entirely at low densities ($n < 10^9$ cm$^{-3}$)

This disk and envelope evolution resembles that in \cite{stacyetal2010}, particularly in that early gravitational heating by the sink caused the development of a phase of hot and dense gas.  However, the hot phase in \cite{stacyetal2010} fell mostly between densities of $10^8$ and $10^{12}$ cm$^{-3}$, and the mass of both the envelope and total star-disk system showed a steady increase throughout the simulation.  In contrast, in our current `no-feedback' case, the hot phase lies mostly between densities of  $10^6$ and $10^9$ cm$^{-3}$, and both the disk and envelope decline in mass after 2000 yr.  This difference between simulations ultimately arises from the statistical variation in sink N-body dynamics,
and the corresponding response from the surrounding gas.

\subsubsection{With-feedback case}
\begin{figure*}
\includegraphics[width=.9\textwidth]{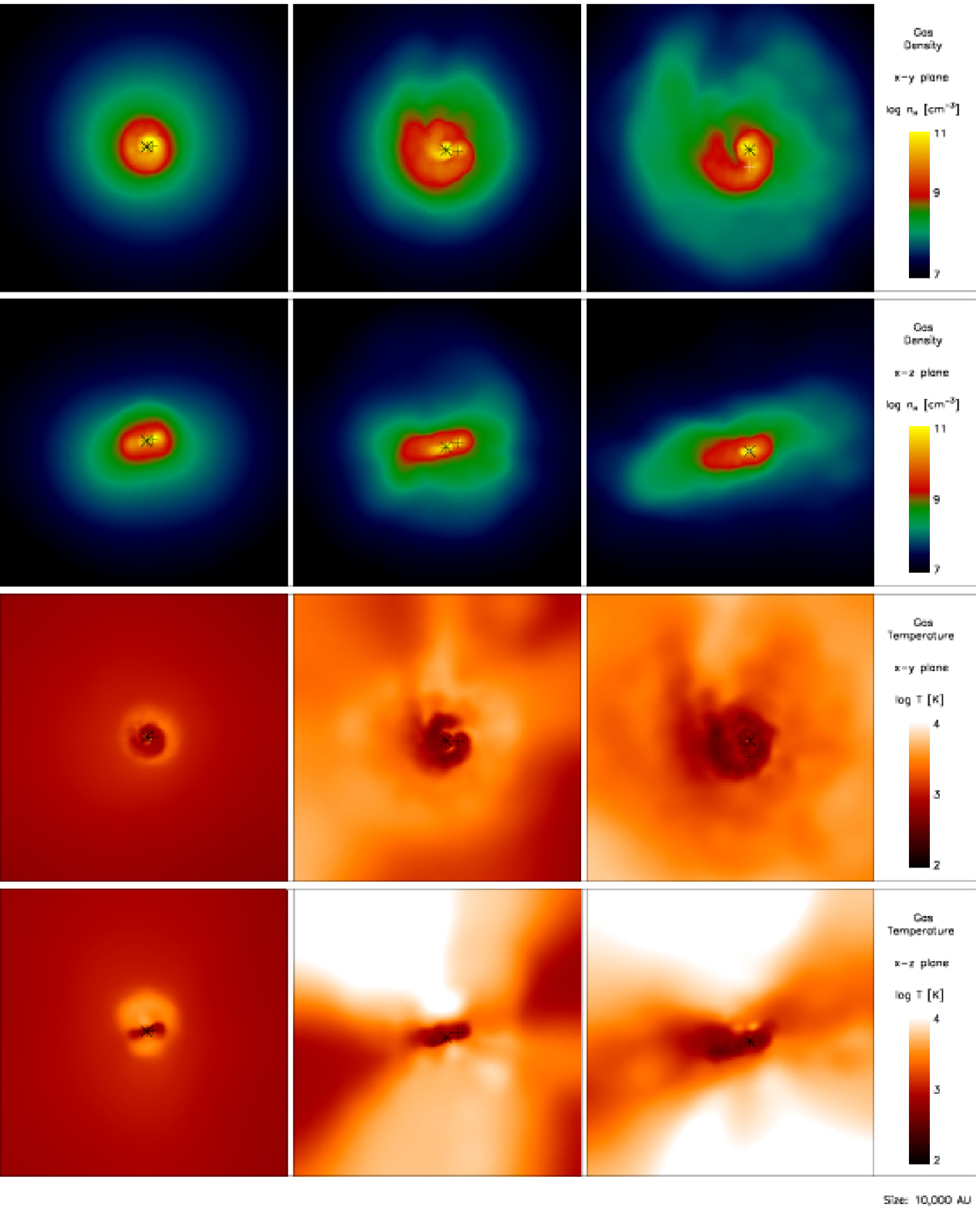}
\caption{Projected density and temperature structure of gas under LW and ionization feedback at 1000, 2000, and 3000 yr after 
initial sink formation.   Asterisks denote the location of the most massive sink.
Crosses show the location of the second-most massive sink.  Diamonds are the locations of the other sinks.
For the rows and columns, we adopt the same convention as employed in Fig.~\ref{morph_nf}.  Note the growth of a roughly hour-glass shaped structure of hot gas surrounding the disk as the I-front expands into the low-density regions.  The hot region expands well beyond the disk by 
 2000 yr.}
\label{morph_nr}
\end{figure*}

The `with-feedback' case also exhibits a peak in $\chi_{\rm rad}$ coincident with the period of rapid sink formation, with a maximum of three sinks in the disk at any given time. This causes the disk to be dominated by N-body dynamics and disrupts the rotational structure, but without ejection of any sink.  However, the disk growth is soon slowed by a different mechanism than that in the `no-feedback' case.  Just as in the latter, we also note a smooth early growth in the mass of hot dense gas (Fig. \ref{diskmass}), initially sourced by the gravitational potential of the main sink.  Once the sink is large enough to emit 
 significant amounts of LW and ionizing radiation, 
after $\sim$ 900 yr of accretion, infalling mass that would otherwise be incorporated into the disk is instead heated to become a hot neutral shell of several thousand Kelvin enclosing a smaller ionized bubble around the disk (see bottom panels of Fig. \ref{Tvsnh} and Fig. \ref{morph_nr}). This leads to the continued increase in the mass of the hot dense gas apparent in Figures \ref{diskmass} and \ref{morph_nr} beyond that seen in the `no-feedback' case.  As the ionization front and hot neutral region expand and rarefy the gas, both the disk and hot dense regions lose mass.  This is visible as a drop in mass of the hot dense gas at 1500 yr 
in Figure \ref{diskmass}, once the pressure wave has reached beyond the dense region into  $n <$ 10$^9$ cm$^{-3}$ gas. 
Throughout this early disk evolution, the total mass of the outer envelope steadily increases (blue line in Fig. \ref{diskmass}) until the same pressure wave passes through the edge of the envelope after 2000 yr, 
after which both the envelope and disk gradually lose mass until the end of the simulation.

 At the same time, angular momentum causes the disk to expand in radius, making the disk more diffuse (see panel {\it a} of Fig. \ref{disk}).  
Once the hot pressure wave travels into the outer envelope, the remaining disk mass is able to level off in density after 2000 yr.  Meanwhile, 
the shielded regions of the disk remain steady in their rotational structure (panels {\it b} and {\it d} of Figure \ref{disk}).  It is interesting to note that, because of disk shielding, for the first 4000 yr N-body dynamics had a greater effect on the disk growth in the `no-feedback' case than protostellar radiation in the `with-feedback' case.  However, the scale of radiative feedback is much larger than that of N-body dynamics, as can be seen by the greater decline in mass of the outer envelope in the `with-feedback' case.

\subsection{Fragmentation}
Consistent with \cite{smithetal2011}, the accretion luminosity does not heat the disk sufficiently to prevent fragmentation.  Radiative feedback does slightly lower the overall fragmentation rate, however, as a total of eight sinks are formed in the `no-feedback' case versus five sinks in the `with-feedback' case. 
In both cases the second sink forms only $\sim$ 100 yr after the initial sink, but, similar to most of the sinks formed, it is quickly lost to a merger.  The next sinks to form and survive to the end of the simulation are created at 300 and 200 yr in the `no-feedback' and `with-feedback' cases, respectively.  This very quick fragmentation is similar to what was described in Clark et al (2011b), \nocite{clarketal2011b} \cite{greifetal2011}, and \cite{smithetal2011}.  
The last sink forms as late as 2000 yr in the `no-feedback' simulation (see Table \ref{tab1}). 

To understand the formation of the initial disk gravitational instability, we first check to see if the disk satisfies the Toomre criterion:

\begin{equation}
Q = \frac{c_{\rm s} \kappa}{\pi G \Sigma} < 1  \mbox{\ ,}
\end{equation}

\noindent where $c_{\rm s}$ is the soundspeed, $\Sigma$ the disk surface density, and $\kappa$ the epicyclic frequency, which for Keplerian rotation is equal to the disk angular velocity $\Omega$.   For the first few 100 yr both disks have average temperatures of around 1000 K, so $c_{\rm s} \sim$ 2 km s$^{-1}$.  
 $M_{\rm disk} \simeq 25$ M$_{\odot}$, 
and the disk radius is nearly 1000 AU, yielding a disk surface density of 
 $\Sigma$~$\simeq$~$M_{\rm disk}/$$\pi R_{\rm disk}^2$~$\sim$~70~g cm$^{-2}$.  
We also approximate $\kappa$ to be $3\times 10^{-11}$~s$^{-1}$, so $Q \sim$~0.4, satisfying the Toomre criterion.  

The ability of the disk to reach Toomre instability is aided by the rapid cooling time of its gas.  The criterion for fragmentation described in \cite{gammie2001} is $t_{\rm cool} \la 3\Omega^{-1}$, where 
$t_{\rm cool}$ is the cooling time of the disk gas.
We find that the typical value of  $\tau_{\rm cool}$  is $ \sim$ 50 yr, while $3\Omega^{-1} \sim 3000$ yr, easily satisfying this criterion, also referred to as the `Viscous Criterion' by  \cite{kratter&murray2011}.

Figures \ref{morph_nf} and \ref{morph_nr} show the density and temperature morphology within the central 10,000 AU.  The multiple sinks and clumpy disk structure are easily visible here, as expected from the low Toomre parameter, though the particular shape of the disk in each case is very different.  The `no-feedback' case has a clearly visible bifurcated temperature structure, with a cool disk and surrounded by warm gas.  The `with-feedback' also has cool disk gas, but the central region is much more dominated by an hour-glass shaped bubble of hot gas.  

\begin{table}
\begin{tabular}[width=.45\textwidth]{crrrr}
\hline
sink  & $t_{\rm form}$ [yr] & $M_{\rm final}$ [M$_{\odot}$]  & $r_{\rm init}$ [AU] & $r_{\rm final}$ [AU]\\
\hline
1  & 0  & 27  & 0 & 0\\
2  & 300 &  0.9 & 100 & 2330\\
3  & 2000  & 2.75 & 70 & 83\\
\hline
\end{tabular}
\caption{Formation times, final masses, distances from the main sink upon initial formation, and distances from the main sink at the final simulation output in the `no-feedback' case. We include the sinks still present at the
end of the simulation (5000 yr).}
\label{tab1}
\end{table}

\begin{table}
\begin{tabular}[width=.45\textwidth]{crrrr}
\hline
sink  & $t_{\rm form}$ [yr] & $M_{\rm final}$ [M$_{\odot}$]  & $r_{\rm init}$ [AU] & $r_{\rm final}$ [AU]\\
\hline
1  & 0  & 19  & 0 & 0\\
2  & 200 &  9.4 & 110 & 440\\
\hline
\end{tabular}
\caption{Same as Table 1, but for sinks remaining at the end of the `with-feedback' case at 4500 yr.}
\label{tab2}
\end{table}

As discussed in \cite{kratteretal2010}, the actual minimum value of $Q$ can vary with disk properties such as scale height, and from Figures \ref{morph_nf} and \ref{morph_nr} we see this varies throughout the disk evolution for both cases.   Thus, it is also necessary to consider other disk properties, particularly the infall rate of mass onto the disk.  The numerical experiments of \cite{kratteretal2010} determined that further fragmentation will occur if the mass infall rate onto the disk is sufficiently high, $\dot{M}_{\rm in} \ga c_{\rm s}^3/G$, such that the disk can no longer process the new material quickly enough.  The star-disk system in the `no-feedback' case grows at approximately  $5 \times 10^{-3}\rm M_{\odot} yr^{-1}$ over the first 2000 yr of infall (Fig. \ref{diskmass}). For the `with-feedback' case the star-disk system grows at this same rate, but 
for a few hundred years longer.  
Comparing this to the average gas soundspeed in the disks,  $\sim$ 2 km s$^{-1}$ in both cases, we find that  $\dot{M}_{\rm in}$ is almost three times greater than $c_{\rm s}^3/G$.  Thus, as expected, the disks in both cases fragment during their initial growth phases, though this phase 
is slightly more unstable for the cooler gas of the `no-feedback' case (panel {\it c} of Figure \ref{disk}), 
for which the formation of the last sink also coincides with a final `bump' in the disk mass 
 and density 
at 2000 yr.  
Though there is a similar bump in disk mass at this time for the `with-feedback' case, the gas is a few hundred Kelvin warmer.  This, along with the steadily decreasing average density of the disk (panel {\it a} of Figure \ref{disk}), is sufficient to keep the  disk stable at later times. 
No new sinks form once the growth of the disks is halted.

Once the disk has become unstable to fragmentation, 
 the continued gravitational collapse of the disk fragments 
furthermore requires that the gas cool quickly enough to overcome opposing effects such as pressure and tidal forces (the `Stalling Criterion'), as described in the recent work of \cite{kratter&murray2011}.  They find that this criterion should be easily met for molecular gas ($\gamma = 7/5$), which applies to the densest regions of the disks in our simulations.  As mentioned above, for the first few hundred yr after sink formation, the average cooling time of the disk is $\tau_{\rm cool} \sim$ 50 yr.  We compare this to the $\beta$ factor described in \cite{kratter&murray2011} to find $\beta = \tau_{\rm cool} \Omega < 0.1$.  The critical value of $\beta$  necessary for free-fall collapse of the fragment ranges from 2.5 to 13 depending on $\gamma$, and the disks easily satisfy this criterion, thus leading to the early formation of sinks within the disks.  

However, in the densest regions of the disk, the `Collisional Criterion' described in \cite{kratter&murray2011}, which requires that the sinks collapse on roughly the orbital timescale, is harder for these sinks to satisfy.  When the disks are undergoing fragmentation during the first few hundred yr,  the sinks form within $\sim 100$ AU of each other and orbit at approximately 5-10 km s$^{-1}$, yielding an orbital timescale on the order of 100 yr.  This is not significantly longer than the free-fall timescale of the sinks, which is $\la 100$ yr.  This leads to sink merging before an entire orbit is complete, particularly during early times in the `no-feedback' case.  The one merger that occurs for the main sink in the `with-feedback' case occurs after multiple orbits, however, and is a result of migration.

\begin{figure*}
\includegraphics[width=.9\textwidth]{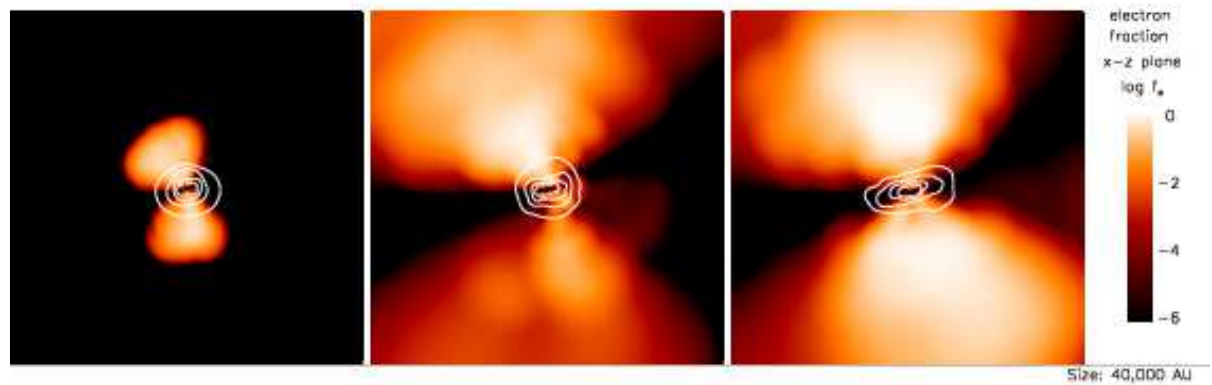}
\caption{Projected ionization structure of gas at 1500, 2000, and 3000 yr after initial sink formation.  White lines depict the density contours of the disk at densities of $10^{7.5}$,  $10^{8}$,  $10^{8.5}$, and  $10^{9}$ cm$^{-3}$.  Box length is 40,000 AU.  Note the pronounced hour-glass morphology of the developing ultra-compact H~{\sc ii} region, roughly perpendicular to the disk.  
This structure gradually expands and dissipates the disk gas from above and below, reducing the scale height of the disk.} 
\label{morph_ifront}
\end{figure*}

\begin{figure*}
\includegraphics[width=.9\textwidth]{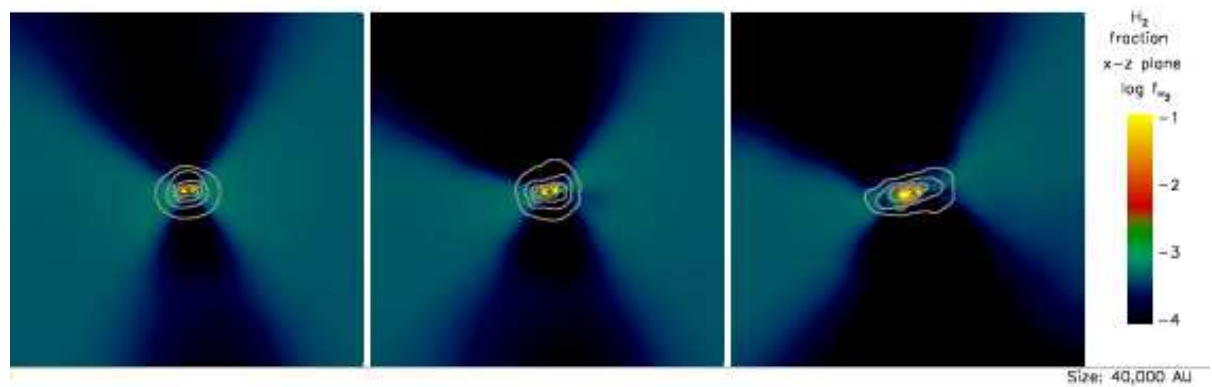}
\caption{Projected H$_2$ fraction of gas at 1500, 2000, and 3000 yr after initial sink formation.  Gray lines depict the density contours of the disk at densities of $10^{7.5}$,  $10^{8}$,  $10^{8.5}$, and  $10^{9}$ cm$^{-3}$.  Box length is 40,000 AU.  Note how the morphology of the neutral region is similar to that of the ionized region.  The molecular inflow gradually becomes confined to the disk plane. }
\label{morph_h2}
\end{figure*}

\subsection{Evolution of Ionization Front}

\begin{figure*}
\includegraphics[width=.8\textwidth]{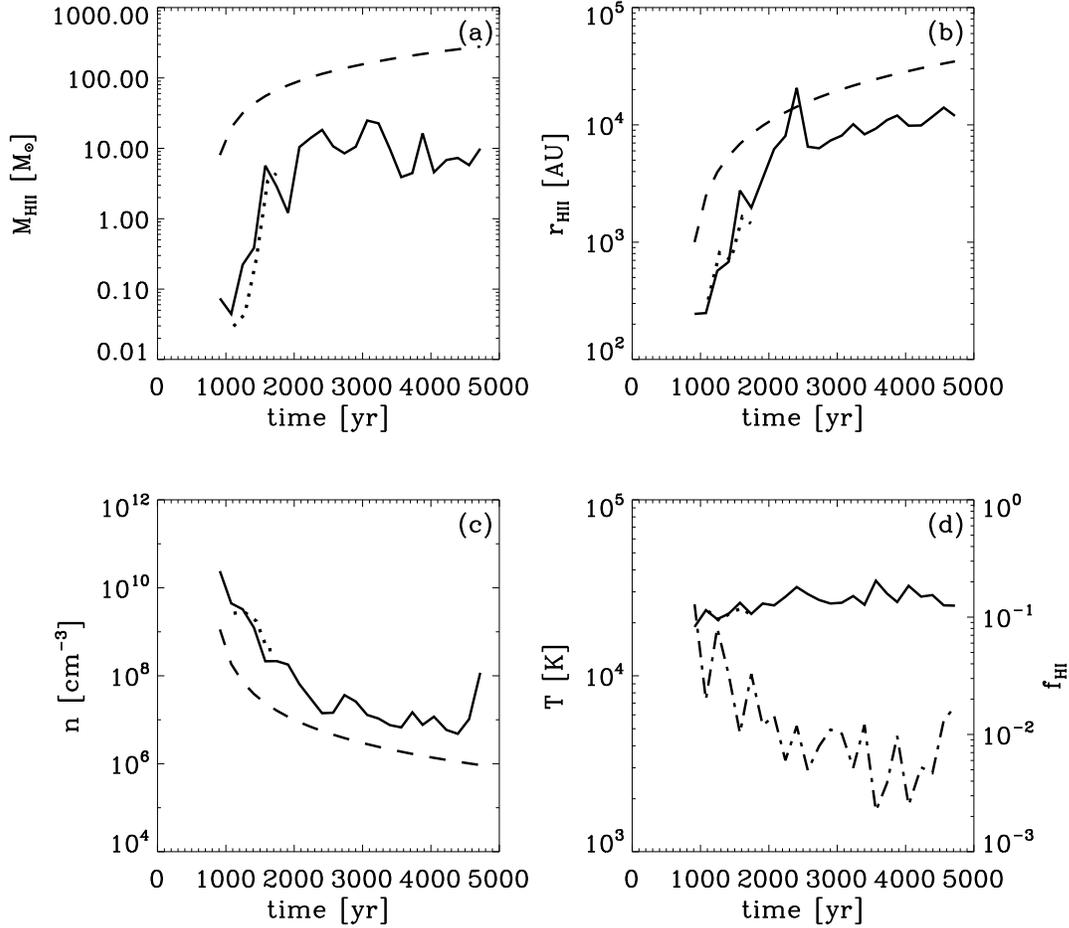}
\caption{Evolution of various H~{\sc ii} region properties over time.  Solid line is taken from the simulation, while the dashed line is the prediction of the self-similar Shu champagne flow solution. 
Dotted line also shows the I-front evolution found from the test simulation in which the ray-tracing was updated at each timestep instead of every five timesteps.
{\it (a):} Total mass of the H~{\sc ii} region.  
{\it (b):} Radial extent of the H~{\sc ii} region, taken as the average distance between the sink and the ionized particles. 
{\it (c):} Average density of the ionized particles.
{\it (d):} Average temperature of the ionized particles (solid line).  
Also shown is the average abundance of neutral hydrogen, $f_{\rm HI}$ (dash-dotted line).
Note how the H~{\sc ii} region fluctuates on small timescales while the long-term evolution generally follows the predicted analytical solution.}
\label{if_evol}
\end{figure*}

The I-front initially appears around the main sink $\sim$ 1000 yr after the sink first forms, once it has grown to $\simeq$ 15 $\rm M_{\odot}$ (see panel {\it d} of Fig. \ref{star-model}).    The morphology of the growing I-front is shown in Figure \ref{morph_ifront}, where we can see that the I-front expands as an hour-glass shape above and below the disk. 
 This same morphology describes the growing neutral region as well (Fig. \ref{morph_h2}), which we will discuss  in the next section (3.4). 
As the I-front expands and widens to encompass a larger angular region, it dissipates the more diffuse disk gas above and below the mid-plane, leading to a decline in the disk scale height.  
Figure \ref{if_evol} compares the I-front evolution with that predicted by the analytical Shu champagne flow solution (\citealt{shuetal2002}, see also \citealt{alvarezetal2006}). Different analytical solutions can be found for the evolution of gas under the propagation of an I-front into a powerlaw density profile, which in our case is approximately $\rho \propto r^{-2}$.  The ratio of the un-ionized gas temperature to that of the H~{\sc ii} region must also be specified for the analytical solution, and in our case we choose a ratio of 20,000 K to 1000 K, or 0.05.  The propagation of the I-front can then be described, assuming D-type conditions and that the I-front closely follows the preceding shock, by a velocity of

\begin{equation}
v_s = x_s c_{\rm s} \mbox{,}
\end{equation}

\noindent while the size of the I-front is

\begin{equation}
r_s = x_s c_{\rm s} t   \mbox{,}
\end{equation}

\noindent where $c_{\rm s}$ is the ionized gas soundspeed and $x_s$ is the position of the shock in similarity coordinates, which for our case is $x_s = 2.54$ (see panel {\it b} of Figure \ref{if_evol}).  

Note that the typical neutral hydrogen abundance $f_{\rm HI}$ is around 10$^{-2}$ in the ionized region, while typical densities in the latter part of the simulation are 10$^7$ cm$^{-3}$  (panels {\it c} and {\it d} of Fig. \ref{if_evol}).  For a hydrogen photoionization cross section of  
$\sigma_{\rm ion} = 6 \times 10^{-18}$ cm$^2$, 
this yields a mean-free-path of

\begin{equation}
l_{\rm mfp} = \frac{1}{\sigma_{\rm ion}n_{\rm HI}} \simeq  \frac{2 \times 10^{17} {\rm cm}}{f_{\rm HI} \, n  \rm [cm^{-3}]} \simeq 2 \times 10^{13} \rm cm \mbox{,}
\end{equation}
 
\noindent where $n_{\rm HI}$ is the number density of neutral hydrogen.  The value of $l_{\rm mfp}$ will usually be $\sim$ 1 AU.  This is much smaller than the typical length of a radial bin in the simulation, with the exception of the small 1.5 AU bins within 75 AU of the star, and also much smaller than the resolution length of the simulation.  Recombination photons should therefore be reabsorbed before exiting their ray-tracing bin and entering neighboring bins.
Our ray-tracing scheme thus loses little accuracy by applying the `on-the-spot' approximation and ignoring the effects of diffuse radiation within the H~{\sc ii} region.

Multiple factors cause deviations from the analytical solution.  For instance, the density structure of the gas is not spherically symmetric.  Furthermore, the initially close proximity of the ionization front to the sink causes the gravity of the sink to have a non-negligible effect on the early H~{\sc ii} region dynamics, a factor that is neglected in the analytical solution.  However, this effect loses importance as the H~{\sc ii} region grows well beyond the gravitational radius, $r_{\rm g} \simeq G M_*/c_{\rm s} \simeq 50$ AU for a $15 M_{\odot}$ star and 20,000K H~{\sc ii} region.  The analytical solution also does not take into account the continued infall of gas onto the central region.  The outer envelope beyond the disk, which we take as the gas with density greater than 10$^8$ cm$^{-3}$, grows for the first 2000 yr at a rate of $\simeq 10^{-2}$  M$_{\odot}$ yr$^{-1}$, or $5 \times 10^{47}$ neutral particles per second.  In comparison, the sink  
typically emits $\ga$ $10^{48}$
photons per second.  Though the infall is not large enough to quench the H~{\sc ii} region entirely, it is a large enough fraction of the ionization rate to cause the total mass of the H~{\sc ii} region to fluctuate.  The motion of the sink through the disk causes the immediately surrounding density structure to vary as well, also contributing to the H~{\sc ii} region's unsteady growth.  
Finally, we note that the peak in the I-front radius at $\sim$ 2500 yr (panel {\it b} of Figure \ref{if_evol}) corresponds to an increase in the luminosity of the main sink due to a concurrent enhancement of the accretion rate. 

This fluctuation is similar to what \cite{galvanetal2011} described in their numerical study of hypercompact H~{\sc ii} regions.  However, our I-front evolution differs significantly from the analytical study by \cite{omukai&inutsuka2002}, where they found that a spherical free-falling envelope would not be unbound by an I-front typically until the Pop III star reaches well over 100  M$_{\odot}$.  In our study the gas does become unbound in the regions polar to the disk, thus indicating how variations in three-dimensional structure and accretion flow play a crucial role in a Pop III star's accretion history.

Because we do not resolve the gas on scales smaller than $\sim 50$ AU, there may be unresolved substructure which would have provided shielding and altered the H~{\sc ii} region growth.  Our simulation only approximates this shielding effect with a simple prescription (Section 2.5) that assumes a smooth and dense sub-sink disk that is coplanar with the large-scale disk resolved in our simulation.  The true shielding within the sink may vary from this assumption, however, depending on the details of the sub-sink structure.  

Clark et al. (2011b) and \cite{greifetal2011} find that disk formation indeed continues down to $< 50$ AU scales, though the disk fragments and clumps on these scales instead of maintaining a smooth structure.  
Were sub-sink scales in our simulation to contain a similar clumpy disk structure that was coplanar with our $\sim$ 1000 AU disk, this would still shield much of the gas lying within the disk plane, as provided by our prescription.  However, assuming no change in the available sub-sink disk mass, clumpiness in the disk will likely increase the escape fraction of ionizing radiation from the sink, since extra radiation may leak through `holes' in the disk (see, e.g. \citealt{wood&loeb2000}).  The growth of the H~{\sc ii} region in our simulation would then be an underestimate. 
On the other hand,  clumpiness along lines-of-sight perpendicular to the disk plane would provide extra shielding in these directions and slow the initial growth of the H~{\sc ii} region in the polar directions.  Our calculation would then have overestimated the H~{\sc ii} region size, particularly given the $n^2$ dependence of the recombination rate.  Our shielding prescription is thus a median between these two possibilities. 

As for structure within the H~{\sc ii} region, our prescription ensures that gas is not ionized unless its density is approximately 100 times lower than our  maximum resolvable density ($\sim 10^{10}$ cm$^{-3}$ versus $10^{12}$ cm$^{-3}$).  Once ionized, gas particles may condense to 100 times their initial density before structure within the H~{\sc ii} region becomes unresolved.  Formation of structure on small $\sim$ 50 AU scales within the H~{\sc ii} region does not occur in our calculation, however, and H~{\sc ii} region substructure should not significantly alter its evolution.          

To test whether performing the ray tracing procedure every five timesteps was sufficient to accurately follow the growth of the I-front, we also performed a test simulation in which the ray tracer was updated at every timestep.  The test simulation was initialized using the output of the `with-feedback' simulation at 1000 yr, and then followed for the following $\sim$ 1000 yr.  As seen in panels {\it a} and {\it b} of Figure \ref{if_evol}, increasing the frequency of the ray-tracing update led to little change in the I-front evolution, so updating every five timesteps was sufficient.

\subsection{Evolution of Hot Neutral Region}

Along with the H~{\sc ii} region, suppression of H$_2$ cooling by LW radiation, 
 combined with continued gravitational heating by the sink,
leads to the development of a growing region of hot neutral gas that expands in a bubble around the disk.
There is a pressure wave of warm  gas, sourced by the combination of the sink's gravitational potential and LW radiation, that expands at the soundspeed
of approximately 7 km s$^{-1}$ (see also Section 3.1).  After 3000 yr this pressure wave has expanded to a radial extent of $r_{\rm pres} \sim 5000$ AU, matching well the predicted size of $r_{\rm pres} = c_{\rm s} \, t = 4400$ AU. 
 This corresponds to the region of gas in Figure \ref{Tvsnh} centered around 10$^8$ cm$^{-3}$ and sitting between approximately 3000 and 7000 K.  

This inner pressure wave is not the sole source of hot neutral gas, however. 
The hot neutral region also has an extended `tail' of 7000 K gas that can be seen in Figure \ref{Tvsnh}.  
This corresponds to the extended region that became ionized during the main sink's enhanced accretion phase, but then recombined and dropped to temperatures of 7000 K when the sink's luminosity declined again.  LW radiation prevents this `tail' of hot neutral gas from cooling back down again.  Thus, LW radiation, gravitational heating by the sink, and recombination of ionized gas all contribute to the growth of the hot neutral region that encompasses the I-front. 

We note that the photodissociation front extends even further than this, beyond 10$^5$ AU (nearly a pc) from the sink by the end of the simulation.  The edge of the photodissociation front is comprised of neutral but cool gas.  
This photodissociation region can expand much more quickly than the gas soundspeed, and thus is much larger than the scale encompassed by the pressure wave.  Following \cite{johnsongreif&bromm2007} and \cite{abeletal1997}, we can estimate the dissocation time $t_{\rm diss, H_2}$ of gas at the edge of this region.  The main sink has a LW luminosity of 10$^{48}$ s$^{-1}$, while the average H$_2$ column density in the neutral region is $\sim 3 \times 10^{19}$ cm$^{-2}$.  We then get an expression similar to equation 9 of \cite{johnsongreif&bromm2007}, but differing by a factor of ten because we consider a less luminous star:

\begin{equation}
t_{\rm diss, H_2} \sim 10^6{\rm \,yr} \left(\frac{R}{{\rm 1 kpc}}\right)^2  \left(\frac{N_{\rm H_ 2}}{10^{14} {\rm \,cm^{-2}}}\right)^{0.75} \mbox{\ .}
\end{equation}

\noindent From this we find that gas at a distance of 0.6 pc will indeed have $t_{\rm diss, H_2} \simeq 5000$ yr, the time of our simulation.   Disk shielding leads to a morphology of the neutral region quite similar to that of the ionized region, as can be seen in Figure \ref{morph_h2}, which shows the expansion of the neutral region as molecular inflow becomes more confined to the disk plane.

\subsection{Protostellar Mass Growth}

The sink growth in both cases is very similar for the first $\sim$ 200 yr, up to when the main sinks reach 12 M$_{\odot}$.  This is similar to some of the simulations presented in \cite{smithetal2011} in which feedback did not cause a significant decline in sink growth until it grew to $> 10$ M$_{\odot}$.  Afterwards, we find the deviation in sink accretion history to be significant, leading to a final mass of 27 M$_{\odot}$ in the `no-feedback' case and 
20 M$_{\odot}$ 
in the `with-feedback' case.  

\begin{figure}
\includegraphics[width=.45\textwidth]{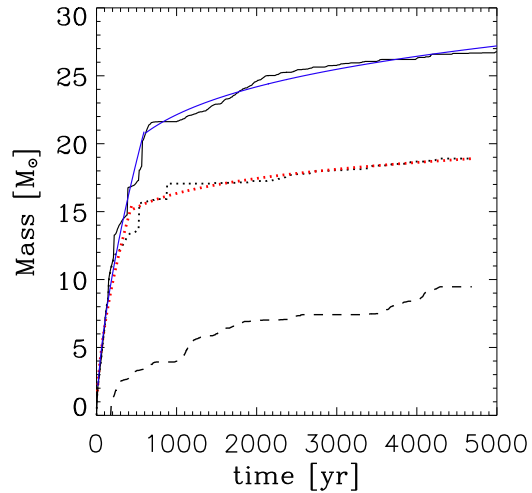}
\caption{Effect of radiative feedback on protostellar accretion.  Black solid line shows mass growth with no radiative feedback, while black dotted line shows the `with-feedback' case.  The blue solid line is a double powerlaw fit to the sink growth rate for the `no-feedback' case, and the red dotted line is a double powerlaw fit for `with-feedback' case.   
The dashed line shows the growth of the second-most massive sink in the `with-feedback' case.
Radiative feedback along with fragmentation-induced starvation leads to a divergence in the accretion histories in less than 1000 yr, and in the `with-feedback' case the main sink does not grow beyond $\sim$ 20 M$_{\odot}$ in the time of the simulation. }
\label{sinkmass}
\end{figure}

\begin{figure}
\includegraphics[width=.45\textwidth]{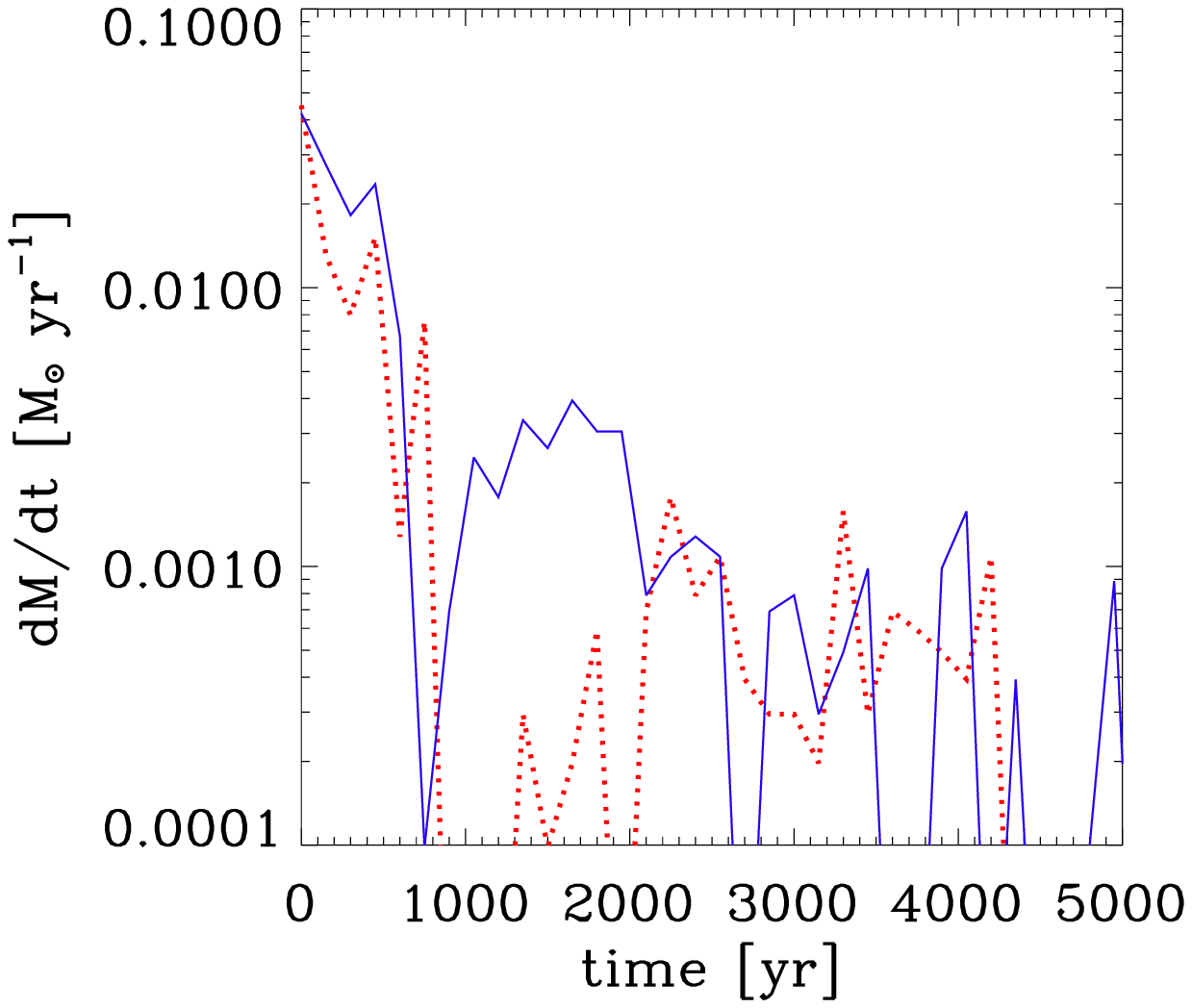}
\includegraphics[width=.45\textwidth]{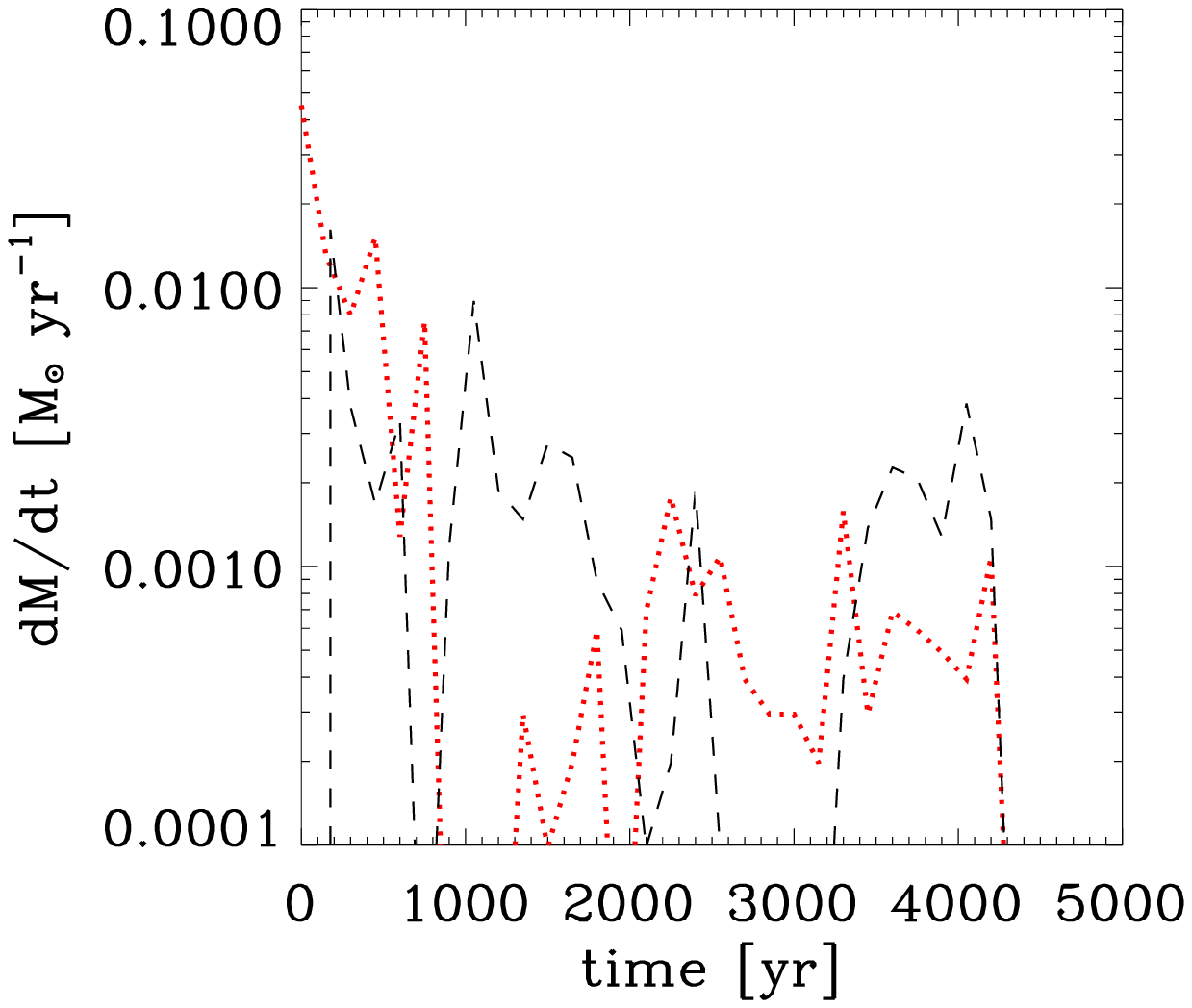}
\caption{
Accretion rate of various sinks in the simulations.  Red dotted line, shown in both the top and bottom panels, represents the main sink for the `with-feedback' case.    Blue solid line (top panel) represents the main sink of the `no-feedback' case.  Black dashed line (bottom panel) is for the second-largest sink of the `with-feedback' case.  Note the generally larger accretion rate of the `no-feedback' case as compared to the accretion rate under feedback.  After 1000 yr, the accretion rates of the two largest sinks in the `with-feedback' case generally alternate in terms of which has the larger amplitude.  The growth of the main sink is especially low between 1000 and 2000 yr, a result of fragmentation-induced starvation.
}
\label{mdot}
\end{figure}

In our `no-feedback' case, the average accretion rate is $5.4 \times 10^{-3}$  $\rm M_{\odot}\, yr^{-1}$ (Fig. \ref{mdot}).  For the first 600 yr,  $M_{*}$ evolves with time as $t_{\rm acc}^{0.64}$.  After the ejection of the secondary sink, the growth rate declines significantly to  $M_{*} \propto t_{\rm acc}^{0.13}$ (see Figure \ref{sinkmass}).  Even after the growth of the disk and envelope is halted, tidal torques are able to continuously funnel additional mass onto the sink.

With feedback, the average accretion rate over the entire simulation drops to  
$\rm 4.2 \times 10^{-3} M_{\odot}\, yr^{-1}$.  
The sink grows more slowly as $M_{*} \propto t_{\rm acc}^{0.56}$ for the first 500 years, and then $M_{*} \propto t_{\rm acc}^{0.09}$ afterwards.  To describe the mass growth history in a different way, we also find that the average accretion rate declines from $3 \times 10^{-2}$  $\rm M_{\odot}\, yr^{-1}$ over the first 500 years to 
 $7 \times 10^{-4}$ M$_{\odot}$ yr$^{-1}$ over the remainder of the simulation.  
Given this strong protostellar feedback onto the disk, we can devise a simple analytical estimate for the final mass attainable by Pop III stars as follows:

\begin{equation}
M_* = \dot{M}_{*} t_{\rm feed}   \mbox{,}
\end{equation}

\noindent where, $ t_{\rm feed}$ is the timescale over which Pop III accretion will be essentially shut off.  Given that Pop III stars must accrete through a disk, and assuming for simplicity the thin-disk approximation,

\begin{equation}
\dot{M}_{*} = 3 \pi \Sigma \nu   \mbox{,}
\end{equation}

\noindent where $\nu$ is the effective viscosity within the disk.  We can approximate this using

\begin{equation}
\nu = \alpha c_{\rm s}^2 / \Omega   \mbox{,}
\end{equation}

\noindent where $\alpha$ is the disk viscosity parameter (\citealt{shakura&sunyaev1973}).  For gravitational torques within strongly self-gravitating disks, $\alpha \propto \left(t_{\rm cool}\Omega\right)^{-1}$ and typically ranges from 0.1-1 (e.g. \citealt{lodato&rice2005}).  In cases like the central regions of our disk closest to the main sink, where $t_{\rm cool} \sim \Omega^{-1}$ (see Section 3.2), the value for $\alpha$ should be closer to 1. 

We now have the expression

\begin{equation}
M_* = 3 \pi \Sigma c_{\rm s}^2 \frac{\alpha}{\Omega} t_{\rm feed}   \mbox{.}
\end{equation}

\noindent If we estimate that the feedback begins once the 
  pressure wave of warm neutral gas 
reaches the scale of the disk radius, which will occur over the sound-crossing time $t_{\rm sound}$, we then have $t_{\rm feed} \sim t_{\rm sound} \simeq 1000$ AU / 7 km s$^{-1}$ $ \simeq 700$ yr. Using  $\Sigma = 140$ g cm$^{-2}$, $\alpha = 1$, $\Omega = 3\times 10^{-11}$~s$^{-1}$, and $c_{\rm s}=2$ km s$^{-1}$, we have a final mass of $M_* = 20$  M$_{\odot}$, well approximating the final protostellar mass reached in our `with-feedback' case. 

Given the rapid evolution  
the disk in both cases at later times, let us 
compare our simulations with analytical estimates of how the sink accretion rate should decline. 
In the `no-feedback' case, at 1000 yr the disk mass has dropped to slightly more than half of its value at the time of initial sink formation, and the mass continues to decline further to one-fourth the intial value.  If we also account for an increase in disk size due to angular momentum conservation, then $\Sigma$ decreases by nearly an order of magnitude. The disk temperature has also dropped from 1000 to 500 K by the end of the simulation (see Figure \ref{disk}), leading to a decrease in $c_{\rm s}$.  From this we approximate that the sink accretion 
rate should again drop by over an order of magnitude after 1000 years, which indeed occurs for the `no-feedback' case.
 
In the `with-feedback' case, by 3000 years of sink accretion the mass of gas in the disk has declined from 40 M$_{\odot}$ to 
$\sim$ 20 M$_{\odot}$, a factor of two.  
$\Sigma$ thus decreases by the same amount.  The disk temperature also decreases from 1000 to 600 K.  Considering equations (25) and (26) 
along with the further decline of $\Sigma$ due to expansion of the disk,
this implies that the accretion rate should decrease by an order of magnitude, from $2 \times 10^{-2}$ M$_{\odot}$ yr$^{-1}$ over the first 1000 years to $\sim 2 \times 10^{-3}$ M$_{\odot}$ yr$^{-1}$ afterwards.  
While the main sink accretes almost an order of magnitude more slowly than this after 1000 yr ($\sim 4 \times 10^{-4}$ M$_{\odot}$ yr$^{-1}$), during this same time the secondary sink accretes at $10^{-3}$ M$_{\odot}$ yr$^{-1}$ (Fig. \ref{mdot}).  The total mass of the sinks thus grows at nearly the expected rate, but the main sink undergoes `fragmentation-induced starvation' as the secondary sink intercepts the majority of the disk inflow, particularly between 1000 and 2000 yr (Peters et al. 2010a\nocite{petersetal2010a}).  After 2000 yr, however, the distance between the sinks increases from $\sim 200$ AU to greater than $\sim 400$ AU, and the secondary sink no longer intercepts the inflowing mass.  
In fact, between 2500 and 3200 yr, the opposite happens and mass flow onto the secondary sink is temporarily intercepted by the main sink.

Also after 2000 yr, radiative feedback begins to shut off mass flow onto the disk, and mass flow onto both of the sinks gradually declines as well.  We can extrapolate the mass growth of the main sink to later times by using the powerlaw fit to the sink's mass after 2000 yr, $M_{*} \propto t_{\rm acc}^{0.12}$.  KH contraction will likely end between roughly 10$^4$ and 10$^5$ yr, and at 10$^5$ yr our fit yields a mass of $\sim 30$ M$_{\odot}$ for the largest star.  For the secondary sink, a fit of $M_{*} \propto t_{\rm acc}^{0.39}$ to the late-time growth also yields a mass of $\sim 30$ M$_{\odot}$ at 10$^5$ yr, implying the stellar system may evolve to an equal-mass binary.  We again point out that we underestimate the Pop III mass, however, due to the early use of ZAMS values in our protostellar model and incomplete disk shielding on small scales.  In comparison, the same extrapolation for the `no-feedback' case gives an asymptotic mass of $\sim 35$ M$_{\odot}$, indicating that N-body dynamics can also play a role in reducing the final Pop III mass. 

We here remind the reader that our sink algorithm allowed for sinks to merge.  The stars represented by the sinks may have instead become a tight binary, and such a binary may have been disrupted by later close encounters with other stars.  Without sink merging, the final mass reached by the main sink could have differed (see also discussion in \citealt{greifetal2011}), and the total stellar content might not have been dominated by a massive binary.  
However, our main sink gained only a small amount of its total mass, $\sim$ 4 M$_{\odot}$, through mergers, and the second-largest sink also gained the majority of its mass through gas accretion.  In the opposite extreme of no sink mergers, a likely outcome may have still been a stellar system dominated by a massive binary, but also including several much smaller sinks.  Determining the actual stellar merger rate of stars that come into close vicinity will be left for future work. 

\section{Summary and Conclusions}

We have performed cosmological simulations of the build-up of a Pop III star to determine how LW and ionizing radiative feedback influences the mass growth of the star and the fragmentation of primordial gas.  We find that radiative feedback will not prevent fragmentation, but will lower the final mass attainable by a Pop III star.  
When accounting for feedback, we estimate a Pop III mass of 30 M$_{\odot}$ by 10$^5$ yr, once it has reached the ZAMS. This is a lower limit, however, due to details in our modeling of the protostellar evolution and inability to fully account for inner-disk shielding.  
Inclusion of feedback also led to a massive binary system instead of a higher-order multiple like that seen in, e.g., \cite{stacyetal2010} and our `no-feedback' case, since the feedback quenched disk growth and fragmentation early on.  In agreement with \cite{smithetal2011}, we furthermore find that stellar N-body dynamics can also play a significant role in the growth of a Pop III star through stellar ejections and disk scattering.  
The final masses reached in our simulation agree well with the recent two-dimensional study by \cite{hosokawaetal2011}, which also examined the effects of radiative feedback on Pop III mass growth and found that accretion was halted at 40 M$_{\odot}$.  

It is interesting to compare our results to that of recent work by Peters et al. (2010b)\nocite{petersetal2010b}.  They similarly examine ionizing and non-ionizing radiative feedback on massive star formation, though they study the case of present-day star formation, and their initial configuration was different from ours in that they began with a 1000 M$_{\odot}$ rotating molecular cloud core.  They find H~{\sc ii} regions which fluctuate in size and shape as gas flows onto the stars, and that the final mass of the largest stars is set by `fragmentation-induced starvation,' a process in which the smaller stars accrete mass flowing through the disk before it is able to reach the most massive star.  This is in contrast to models in which the final stellar mass is set once ionizing radiation shuts off the disk accretion (e.g. \citealt{mckee&tan2008}).  In our `with-feedback' case we find that 
fragmentation-induced starvation occurs between $\sim$ 1000 and 2000 yr, while afterwards radiative feedback does lead to a further decline in the sink accretion rate and the disk mass.  Soon after the I-front breaks out from the sink, the second largest sink accretes much more quickly than the first, intercepting a large portion of infalling mass that otherwise would be accreted by the main sink.  Indeed, the final estimated mass of the main sink is similar to that found in Peters et al. (2010a)\nocite{petersetal2010a}, where they found the combination of disk fragmentation and radiative feedback led to a most-massive star of 25 M$_{\odot}$.  
However, in our case radiative feedback grows in importance after 2000 yr as it eventually slows mass flow onto either sink. 
A similar study of current-day star formation by \cite{krumholzetal2009} found that a prestellar core would similarly collapse into a disk that would host a multiple system of massive stars, even under the effects of radiation pressure.  However, they followed smaller average accretion rates over a longer period of time (50,000 yr) and found that gravitational and Rayleigh-Taylor instabilities would continue to feed mass onto the disk and stars.  

We also note that, despite their very similar numerical setups, the disk evolution and accretion rate of our `no-feedback' case differs somewhat from that described in \cite{stacyetal2010}.  Several distinctions between the simulations, however, explain this.  The high-density cooling and chemistry is updated from that used in \citealt{stacyetal2010} (see Section 2.2).  We also use an adaptive softening length instead of a single softening length for all gas particles as in \cite{stacyetal2010}, and our criteria for sink accretion are slightly more stringent.  The main contribution to the difference, however, is likely the stochastic nature of the sink particle dynamics.  While no sink was ejected in \cite{stacyetal2010}, the sink ejection and subsequent rapid velocity of the main sink in our `no-feedback' case altered the disk structure, and the main sink would likely have grown to a higher mass otherwise.  Nevertheless, the final sink masses in both simulations were still the same to within a factor of two.     

The radiative feedback seen here is much stronger than the analytical prediction of \cite{mckee&tan2008},  
even when considering the combined accretion rate of both sinks.  They found that a Pop III star could grow to over 100 M$_{\odot}$ through disk accretion, as disk shadowing allowed mass to flow onto the star even while the polar regions became ionized.  
Our lack of resolution prevents this disk shadowing 
from being fully modeled on sub-sink scales, and the ionizing photon emission emanating from the sink edge is likely overestimated 
along some lines-of-sight.  This especially applies to angles just above and below the disk that quickly became ionized in our `with-feedback case', when in reality these regions are unlikely to become ionized until sometime later, only after the gas that is polar the disk becomes ionized first. 
\cite{mckee&tan2008} furthermore assumed disk axisymmetry, which does not describe the disk in either of our test cases.    
Nevertheless, our shielding prescription does indeed keep 
the disk from becoming ionized in our `with-feedback' case.  However, the I-front does not expand in a perfectly uniform fashion along the polar directions, as the disk rotates, and the position of the main sink within the disk varies as it orbits its companion sink. Thus, different angles will be shielded at different times.  Once gas along a certain direction has been ionized, it may recombine at later times, but LW radiation prevents most of this gas from cooling back down to below a few thousand Kelvin.  
  LW radiation, combined with recombination of ionized gas and gravitational heating from the sink, therefore leads to a bubble of hot neutral gas that continues to expand in all directions,
except within a few degrees of the disk plane.  Mass flow onto the disk and sinks is then greatly reduced.  Thus, while our results underestimate the effect of shielding, it still highlights how non-axisymmetry will enhance the effects of radiative feedback, and the true physical case likely lies somewhere in between our `with-feedback' case and the prediction of \cite{mckee&tan2008}. In a similar vein, non-axisymmetry can also promote further disk fragmentation, and this in turn can result in N-body dynamics that may provide another means of reducing Pop III accretion rates.     

The fragmentation of primordial gas and growth of Pop~III stars has recently been modeled from cosmological initial conditions with resolution reaching nearly protostellar scales (Clark et al. 2011b, \citealt{greifetal2011}).  However, though our simulation is less highly resolved, it explores a different regime of Pop III growth.  The aforementioned studies could not follow the mass accretion for more than 100-1000 yr, before the protostars had grown beyond 10 M$_{\odot}$, and they did not follow the growth of the I-front.  Our work thus affords a first look, starting from cosmological initial conditions, at how Pop III growth will continue beyond 
1000 yr, after the formation of the I-front.  

In both the `with-feedback' and `no-feedback' cases, the total stellar mass at 1000 yr was similar to the total stellar mass found for the corresponding time in the simulations of \cite{greifetal2011}, which had comparatively more refined resolution, utilized sinks with smaller sizes of $\la 1$ AU, and found fragmentation on scales $<$ 50 AU.  Higher resolution may reveal fragmentation on sub-sink scales in our simulations as well.   Nevertheless, \cite{greifetal2011} found that the highest mass of any individual sink at 1000 yr was already $\sim$10 M$_{\odot}$, large enough to emit substantial radiation.  Our maximum sink masses at this time were slightly larger but still similar, 17 and 22~M$_{\odot}$.  Thus, while our strong feedback yields a lower limit on the total stellar mass of the system, the mass of any given sink is an upper limit to the mass of an individual protostar within that sink.  
Radiative feedback possibly suppresses fragmentation on sub-sink scales, as seen within studies of cold gas hosting present-day star formation (e.g. \citealt{krumholzetal2007}).   Primordial gas is much warmer, however, and \cite{smithetal2011} found only mild suppression of primordial gas fragmentation under non-ionizing feedback from $\sim 10$ M$_{\odot}$ stars.  It remains to be seen whether primordial gas would undergo significantly less sub-sink fragmentation under ionizing feedback from more massive stars.

These remaining issues show that more computationally expensive and highly resolved simulations will be necessary in the future to determine the true nature of sub-sink fragmentation.  
Future cosmological simulations will eventually bridge this gap by both resolving protostellar scales and modeling ionizing radiation, but such a calculation pushes current computational limits.

Even with these caveats, however, our results are largely consistent with observations of massive stars in the Galaxy (e.g. \citealt{masonetal1998,masonetal2009}), showing that about 40\% of O stars have companions in the visual binary regime, corresponding to typical distances of $\sim$ 1000 AU (see also discussion in \citealt{krumholzetal2009}).    
Computational studies of present-day massive star-fomation (e.g. \citealt{krumholzetal2009}) have also found disk structures dominated by a massive binary.    Since similar disk accretion processes are at work in the Pop III regime, our finding of a stellar system dominated by a massive binary is unsurprising.

If radiative feedback were typically able to prevent Pop III stars from growing to more than a few tens of solar masses, this would have several effects on their observational signatures.  PISNe would be less frequent, as this requires a star to grow to greater than 140~M$_{\odot}$.  Old, metal-poor stars within the Milky Way halo and nearby dwarf galaxies may preserve the nucleosynthetic pattern of the first SNe, so this may help to explain the lack of PISNe chemical signatures found in these nearby stars (e.g. \citealt{christliebetal2002,beers&christlieb2005,frebeletal2005,tumlinson2006}; but see \citealt{karlsson2008}). Instead, Pop III stars may end their lives through core-collapse SNe or direct collapse to BHs.  For sufficient stellar rotation rates, the possibility of Pop III collapsar GRBs also remains (e.g. \citealt{stacyetal2011}).  The feedback of Pop III stars on their neighboring metal-free minihaloes would also be altered, though the details of how the mass and formation rate of such `Pop III.2' stars would be affected remains to be determined by future simulations.

\section*{Acknowledgments}

VB acknowledges support from  NSF grant AST-1009928, NASA ATFP grant NNX09AJ33G, and JPL Research Support
Agreement 1354840. The simulations were carried out at the Texas Advanced
Computing Center (TACC).  VB thanks the Max-Planck-Institut f\"{u}r Astrophysik for its hospitality during part of the work on this paper.
The authors would like to thank
Takashi Hosokawa and Naoki Yoshida for valuable comments that helped to
improve the paper. 
The authors would also like to thank the anonymous referee for useful suggestions that enhanced the presentation of this work.


\bibliographystyle{mn2e}
\bibliography{ifront_alt}{}

\label{lastpage}

\end{document}